\documentclass[twocolumn,tighten]{aastex631}

\submitjournal{\apj}

\shortauthors{Afrin \& Ghosh}
\usepackage{graphicx}	
\usepackage{amsmath}	
\usepackage{amsfonts}
\usepackage{float}
\usepackage{textcomp}
\usepackage{wrapfig}
\usepackage{cancel}

\usepackage{eucal}
\usepackage{cancel}
\hypersetup{
    colorlinks=true,
    citecolor=magenta,
    linkcolor=cyan,
}
\begin{document}
\title{Testing Horndeski Gravity from EHT Observational Results for Rotating Black Holes}
\correspondingauthor{Misba Afrin}
\email{me.misba@gmail.com}
\author[0000-0001-5545-3507]{Misba Afrin}
\affiliation{Centre for Theoretical Physics,
Jamia Millia Islamia, New Delhi 110025, India}
\author[0000-0002-0835-3690]{Sushant G. Ghosh}
\affiliation{Centre for Theoretical Physics,
Jamia Millia Islamia, New Delhi 110025, India}
\affiliation{Astrophysics and Cosmology Research Unit, School of Mathematics, Statistics and Computer Science,
University of KwaZulu-Natal, Private Bag 54001, Durban 4000, South Africa}
\begin{abstract}
The Event Horizon Telescope (EHT) collaboration recently unveiled the first image of the supermassive black hole M87*, which exhibited a ring of angular diameter $\theta_{d}=42 \pm 3 \mu as$, a circularity deviation $\Delta C \leq 0.1$, and also inferred a black hole mass of $M=(6.5 \pm 0.7) \times 10^9 M_\odot $. This provides a new window onto tests of theories of gravity in the strong-field regime, including probes of violations of the no-hair theorem. It is widely believed that the Kerr metric describes the astrophysical black holes, as encapsulated in the critical but untested no-hair theorem.  
Modeling Horndeski gravity black holes---with additional hair parameter $h$ besides the mass $M$ and spin $a$ of the Kerr black hole---as the supermassive black hole M87*, we observe that to be a viable astrophysical black hole candidate, the EHT result constrains ($a$, $h$) parameter space. However, a systematic bias analysis indicates rotating Horndeski black hole shadows may or may not capture Kerr black hole shadows, depending
on the parameter values; the latter is the case over a substantial part of the constrained parameter space, allowing Horndeski gravity and general relativity to be distinguishable in the said space, and opening up the possibility of potential modifications to the Kerr metric. 
\end{abstract}
\keywords{Astrophysical black holes (98); Black hole physics (159); Galactic center (565);  Gravitation (661); Gravitational lensing (670)}
\section{Introduction}\label{Intro}
According to the no-hair theorem \citep{Carter:1971zc} of general relativity (GR), black holes are characterized by three gauge charges: mass, spin and electric charge \citep{Israel:1967wq,Hawking:1971vc}. The charge of astrophysical black holes is expected to be negligible \citep{Zajacek:2018ycb} and it is also spontaneously lost in realistic environments \citep{Gibbons:1975kk}; hence, it is a general belief that astrophysical black holes are described by Kerr spacetime. Undoubtedly, GR has served as a very well tested standard model of gravity; nonetheless, modified theories of gravity (MoGs) \citep{Clifton:2011jh} have been actively explored, mainly for quantum field theoretical and cosmological reasons: the pathological occurrence of ghost degrees of freedom due to renormalization of higher-order GR theories \citep{Stelle:1976gc}, the occurrence of singularity \citep{Psaltis:2008bb}, the anomalous acceleration of Pioneer \citep{Anderson:2001sg}, and the experimental evidence suggesting the need for more than 95\% of our Universe to be made from dark matter \citep{Zwicky:1933gu} and dark energy \citep{SDSS:2014irn,AtacamaCosmologyTelescope:2013swu,BOSS:2012dmf}. 

The above, as well as several other recent developments in astrophysics and cosmology beyond GR, have sparked interest in various scalar-tensor theories of gravity. Studies of the universe's mysterious late time acceleration and inflationary phase have called for the coupling of GR to scalar fields \citep{Brito:2019ose}. These efforts led to the development of the well-known generalized Galileons, which can be mapped to the most general scalar-tensor theory in four dimensions, with second-order field equations (of a scalar field $\phi$) and a second-order energy-momentum tensor \citep{Nicolis:2008in}, first proposed in 1974 and termed the Horndeski theory \citep{Horndeski:1974wa}. Besides the cosmological reasons, several exciting investigations in astrophysics lead a fortiori to various black hole solutions in the Horndeski theory: Hawking-Page phase transition in context of asymptotically locally anti de Sitter and flat black holes \citep{Anabalon:2013oea}, constraints from the perihelion precession and the gravitational bending angle of light in spherically symmetric black holes in the Horndeski framework \citep{Bhattacharya:2016naa}, and upper bound on Galelian charge of an exact black hole solution in Horndeski gravity using the the Gravity Probe B results \citep{Mukherjee:2017fqz} are few recent investigations. Furthermore, solutions for compact astrophysical objects have also been considered viz., the construction and analysis of boson stars in the biscalar extension of Horndeski gravity \citep{Brihaye:2016lin} and slowly rotating neutron stars in the nonminimal derivative coupling sector of Horndeski gravity \citep{Cisterna:2016vdx}. 

To test Einstein's GR and ultimately to find out {\em the} correct (effective low-energy and high-energy) description(s) of gravity, we need to know the theoretical predictions of other theories as well. Besides academic interests, this motivation has been ever-increasing following the first detection of gravitational waves~\citep{LIGOScientific:2016aoc} and the first image of the supermassive black hole M87* by the Event Horizon Telescope (EHT) collaboration \citep{Akiyama:2019bqs,Akiyama:2019cqa,Akiyama:2019eap}. As an interferometer, using the Very Long Baseline Interferometry (VLBI) technique, the EHT has recently resolved the central brightness depression in the obtained image, which has been interpreted as the shadow cast by the black hole owing to gravitational lensing of photons originating from the surrounding plasma, with the overall following the expected shadow of a Kerr black hole, as predicted in GR \citep{Akiyama:2019bqs}. The central compact radio source---resolved as an asymmetric bright emission ring---has an angular diameter of $42\pm3\mu$as, wherein the asymmetry arises due to the relativistic beaming of photons. Further, the shadow image is found to exhibit a deviation from circularity, $\Delta C \leq 10\%$ and an axis ratio $\lesssim 4/3$  \citep{Akiyama:2019bqs,Akiyama:2019cqa,Akiyama:2019eap}. 
However, there are several caveats to the predictions of how the M87* would observationally appear, due to the underlying uncertainties. These are associated with the observation itself, due to the different telescopes in the sparse array, as well as the observation being sensitive to many untested accretions and emission physics in the vicinity of the supermassive black hole \citep{Gralla:2020pra}. Nevertheless, subject to the various uncertainties, the EHT observational constraints open up a new way to probe the background metric in the strong-field regime, which we intend to employ to test the viability of black holes in the Horndeski theory, then measure their distinguishability from the Kerr black hole.

We intend to probe the rotating black holes in Horndeski gravity \citep{Walia:2021emv}, by assuming M87* to be one and imposing the astronomical constraints (i) \textit{(i)} $\Delta C \leq 0.1$ and (ii) $39\mu as \leq \theta_d \leq 45\mu as$ on the parameter space ($a$, $h$) of the black holes. Within the constrained parameter space---where M87* can be a rotating Horndeski black hole aw well as a Kerr black hole of GR---a systematic bias analysis is carried out, to find out whether the various shadow observables in Horndeski theory and GR are distinguishable at the current $\sigma=10\%$ uncertainty of the EHT measurements \citep{Akiyama:2019bqs}. The shadow area $A$ and oblateness $D$ are used to define a cost function, the reduced $\chi^2$ over the parameter space ($a$, $h$), to determine whether it is large enough to distinguish the two theories of gravity in question. Thus in principle, the present study would place constraints on the parameters of the rotating black holes in Horndeski's theory and, in turn, will also test the Kerr hypothesis with the EHT observations. 

This paper is organized as follows. In Section~\ref{Sec2}, we inspect the rotating  Horndeski black holes and examine the effect of the $a$ and $h$ parameters on the horizon structure, as well as studying the frame-dragging effect. Section~\ref{Sec3} is devoted to the photon region around the black holes and the impact of the hair parameter $h$ on their shadows, in comparison with Kerr black holes. We characterize the shadows with various observables and use them to estimate the parameters of the rotating Horndeski black holes in Section~\ref{Sec4}. The supermassive black hole M87* is modeled as a rotating Horndeski black hole in Section~\ref{Sec5}, and the parameter space is constrained using the EHT observations. In Section~\ref{Sec6} we carry out a systematic bias analysis to distinguish the Horndeski theory from GR. Finally, in Section~\ref{Sec7}, we summarize our results. 

We use geometrized units $8 \pi G= c =1$, unless the units are specifically defined.
\section{Rotating black holes in Horndeski theory}\label{Sec2}
The Horndeski theory that we consider is a class of scalar tensor theory that involves four arbitrary functions $Q_i(\chi)$ $(i=2,...,5)$, of the kinetic term $\chi =-\partial^\mu \phi \partial_\mu \phi/2$ whose action reads
\begin{eqnarray}
\label{action}
\nonumber
S=\int \sqrt{-g}\Big\{Q_2(\chi)+Q_3(\chi)\square\phi + Q_4(\chi)R\\
+Q_4,_\chi[(\square\phi)^2-(\nabla^\mu\nabla^\nu\phi)(\nabla_\mu\nabla_\nu\phi)]\Big\}d^4x,
\end{eqnarray}
where $g$ is the determinant of the metric and $R$ is the Ricci scalar. The action (\ref{action}) is of a particular type, with $Q_5=0$ \citep{Babichev:2017guv}, for which a static spherically symmetric solution is sought, taking the metric ansatz \citep{Bergliaffa:2021diw} 
\begin{eqnarray}
	\label{non_rot_metric}
	ds^2=-A(r)dt^2+\frac{1}{B(r)}dr^2+r^2(d\theta^2+\sin^2\theta d\varphi^2),
\end{eqnarray}
where $A(r)$ and $B(r)$ are arbitrary functions to be determined. Assuming the $4$-current as $$j^\nu=\frac{1}{\sqrt{-g}}\frac{\delta S}{\delta(\phi_{,\mu})},$$ gives the result \citep{Bergliaffa:2021diw}
\begin{eqnarray}
\label{eQ_2}
\nonumber
j^\nu=-Q_2,_\chi \phi^{, \nu}-Q_3,_\chi (\phi^{, \nu}\square\phi+\chi^{, \nu})~~~~~~~~~~~~~~~~~~~~\\
-Q_4,_\chi (\phi^{, \nu}R-2R^{\nu\sigma}\phi,_\sigma)~~~~~~~~~~~~~~~~~~~~~~~~~~~\\
\nonumber
-Q_4,_\chi,_\chi\{\phi^{, \nu}[(\square \phi)^2
-(\nabla_\alpha\nabla_\beta\phi)(\nabla^\alpha\nabla^\beta\phi)]\\
\nonumber
+2(\chi^{, \nu}\square\phi-\chi,_\mu\nabla^{\mu}\nabla^{\nu}\phi)
\},
\end{eqnarray} 
where the usual convention for the Riemann tensor,   $$\nabla_\rho\nabla_\beta\nabla_\alpha\phi-\nabla_\beta\nabla_\rho\nabla_\alpha\phi=-R^\sigma_{~\alpha\rho\beta}\nabla_\sigma\phi,$$ is used.
\begin{figure}
    \centering
    \includegraphics[scale=0.75]{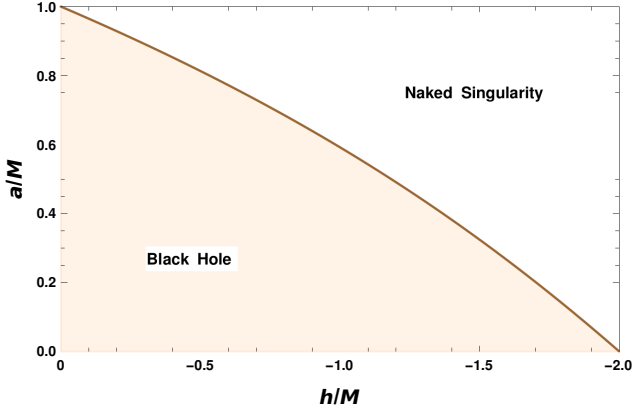}
    \caption{The parameter space ($a$, $h$) for the rotating Horndeski black holes. The solid line separates black holes from naked singularity configurations ($\Delta(r)=0$ has no real roots).}
    \label{parameterSpace}
\end{figure}
Varying the action (\ref{action}) with respect to $g^{\mu\nu}$, the field equations take the form \citep{Bergliaffa:2021diw}
\begin{eqnarray}
\label{eq3}
Q_4 G_{\mu\nu}= T_{\mu\nu},
\end{eqnarray}
where
\begin{eqnarray}
\label{eq4}
\nonumber
T_{\mu\nu}=\frac{1}{2}(Q_2,_\chi \phi ,_\mu \phi ,_\nu+Q_2 g_{\mu\nu})
+\frac{1}{2}Q_3,_\chi(\phi ,_\mu \phi ,_\nu\square\phi~~~\\
\nonumber
-g_{\mu\nu} \chi,_\alpha \phi^{, \alpha}+\chi,_\mu \phi ,_\nu
+\chi,_\nu \phi ,_\mu)- Q_4,_\chi\Big\{\frac{1}{2}g_{\mu\nu}[(\square\phi)^2\\
\nonumber
-(\nabla_\alpha\nabla_\beta\phi)(\nabla^\alpha\nabla^\beta\phi)-2R_{\sigma\gamma}\phi^{,\sigma}\phi^{,\gamma}]
-\nabla_\mu\nabla_\nu \phi \square\phi\\
\nonumber
+\nabla_\gamma\nabla_\mu \phi \nabla^\gamma \nabla_\nu \phi-\frac{1}{2}\phi ,_\mu \phi ,_\nu R 
+R_{\sigma\mu}\phi^{,\sigma}\phi,_{\nu}\\
+R_{\sigma\nu}\phi^{,\sigma}\phi,_{\mu}+R_{\sigma\nu\gamma\mu} \phi^{,\sigma}\phi^{,\gamma}
\Big\}~~~~~~~~~~~~~~\\
\nonumber
-Q_4,_\chi,_\chi \Big\{g_{\mu\nu}(\chi,_{\alpha}\phi^{,\alpha}\square \phi+\chi_{,\alpha} \chi^{, \alpha})+\frac{1}{2}\phi ,_\mu \phi ,_\nu\times
\\
\nonumber
(\nabla_\alpha\nabla_\beta\phi\nabla^\alpha\nabla^\beta\phi-(\square\phi)^2)
- \chi,_\mu \chi,_\nu \\
\nonumber
- \square\phi( \chi,_\mu \phi,_\nu  
+ \chi,_\nu \phi,_\mu)
\\
\nonumber
- \chi,_\gamma[\phi^{,\gamma}\nabla_\mu\nabla_\nu\phi-(\nabla^\gamma\nabla_\mu\phi)\phi,_{\nu}
-(\nabla^\gamma\nabla_\nu\phi)\phi,_{\mu}]
  \Big\}.
\end{eqnarray}
Taking the canonical action for the scalar field  $\phi\equiv\phi(r)$, which is the source of the static and spherically symmetric geometry described by the metric (\ref{non_rot_metric}), imposing conditions of the
finite energy of $\phi$, i.e., $E= \int_V \sqrt{-g}\, T^{0}_{0} \,d^{3}x$ and a vanishing radial $4$-current at infinity $j^{r}=0$, and solving the field Equation (\ref{eq3}), we obtain \citep{Bergliaffa:2021diw}
\begin{eqnarray}
	\label{non_rot_AB}
	A(r)=B(r)=1 -\frac{2M}{r} + \frac{h}{r}\ln\left({\frac{r}{2M}}\right),
\end{eqnarray}
where the integration constant $M$ can be related to the black hole mass and  $h$ is a constant that results from Horndeski theory, referred to as the hair parameter \citep{Bergliaffa:2021diw,Kumar:2021cyl}. The metric (\ref{non_rot_metric}) with (\ref{non_rot_AB}) represents hairy black holes with a scalar polynomial singularity, which always admits a horizon ($r_+=2M$), and thereby respects the cosmic censorship hypothesis \citep{Penrose1999}. A simple root analysis of $B(r)=0$ implies the existence of two positive roots, corresponding to Cauchy and event horizons for $h\in[-2,0]$, whereas, for $ h\in\mathbb{R}\backslash [-2,0]$, the metric has only one horizon; hence we restrict our analysis to the former range. The fact that $\displaystyle{\lim_{ r \to \infty}} {A(r)=B(r)=1}$ guarantees asymptotic flatness. Further, the solution (\ref{non_rot_metric}) with (\ref{non_rot_AB}), in the limit $h\rightarrow 0$, reverts to the Schwarzschild solution. 
\begin{figure*}[t]
\begin{tabular}{c c}
    \hspace{-0.8cm}\includegraphics[scale=0.68]{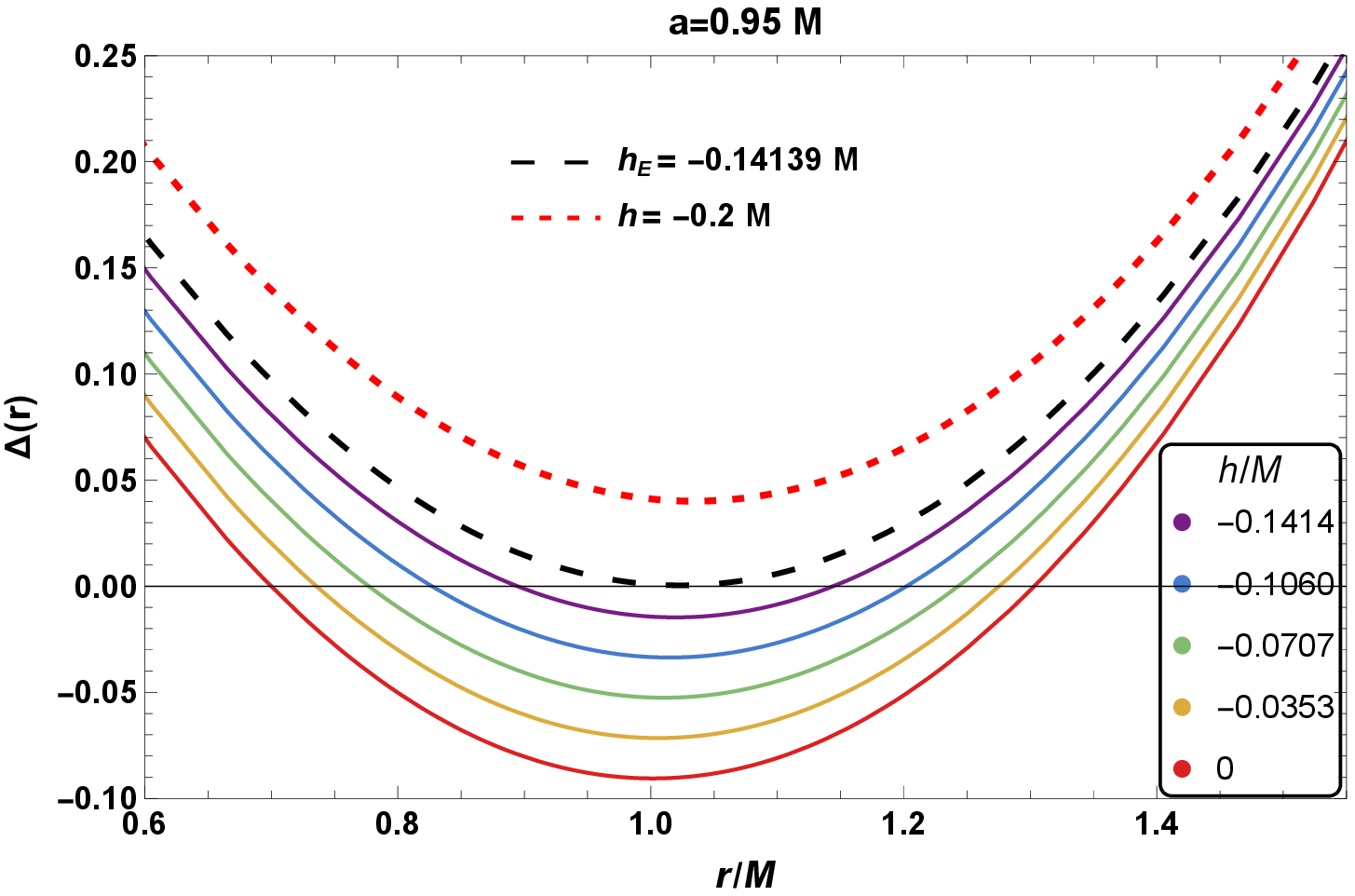}&
    \includegraphics[scale=0.9]{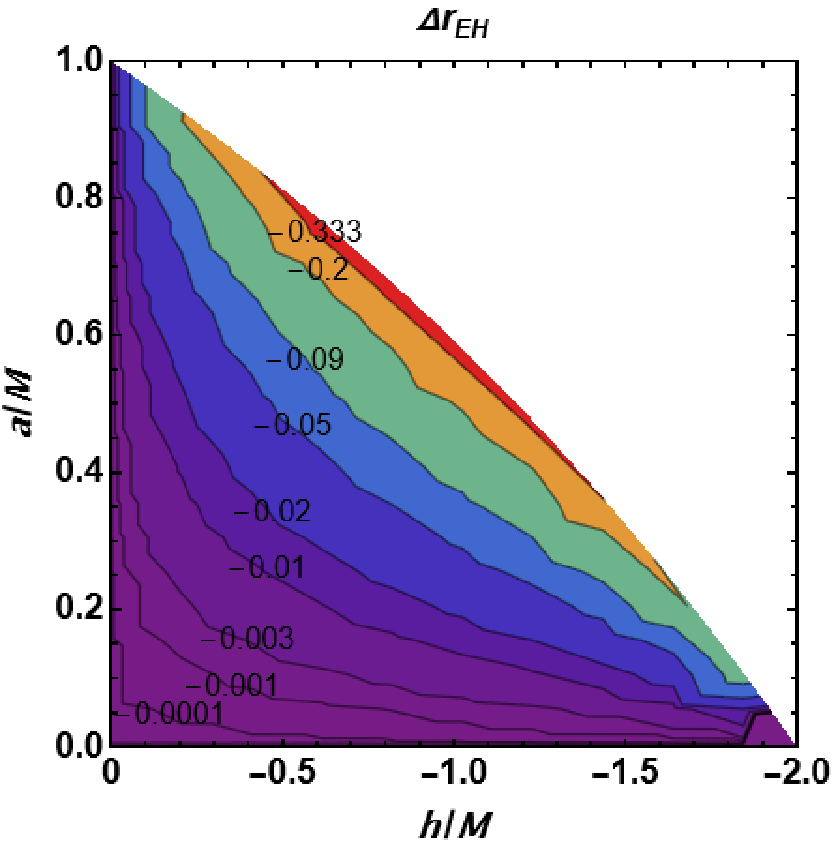}
\end{tabular}
\caption{\label{horizonsFig} Left: horizons of the rotating Horndeski black holes (zeroes of $\Delta(r)=0$) (\textit{left}). Right: constant contours of $\Delta r_{EH}$ as a function of ($a$, $h$) (\textit{right}). $\Delta r_{EH}<0$ implies that the rotating Horndeski black holes have smaller event horizon than the Kerr black holes.}
\end{figure*}
\paragraph{Rotating metric}
The non-rotating black holes cannot be tested by observations, as black hole spin is essential in any astrophysical process. The Kerr metric is one of the crucial GR solutions, representing a rotating black hole that results from gravitational collapse. This prompts us to seek an axisymmetric generalization of the metric (\ref{non_rot_metric}) or find a Kerr-like metric, namely a rotating Horndeski black hole metric, and test it with EHT  observations.  This is achieved via revised Newman-Janis Algorithm (NJA) \citep{Azreg-Ainou:2014pra, Brahma:2020eos}. 
The NJA \citep{Newman1965,Drake:1998gf} has been extensively used to generate rotating black hole metric from the non-rotating counterparts \citep{Johannsen:2011dh}, whereas a revised NJA \citep{Azreg-Ainou:2014pra} has been successfully applied to construct rotating black holes in MoG \citep{Brahma:2020eos}. The rotating counterpart of the black hole (\ref{non_rot_metric}) with (\ref{non_rot_AB}) can be obtained by the revised NJA \citep{Azreg-Ainou:2014pra, Brahma:2020eos} whereby, the rotating black holes in the Horndeski gravity read 
\begin{align}\label{metric2}
ds^{2}&=-\left[1-\frac{2f}{\Sigma}\right]dt^{2}+\frac{\Sigma}{\Delta}dr^{2}-\frac{4af\sin^2{\theta}}{\Sigma}dtd\phi\nonumber\\
&+ \Sigma d\theta^{2}+\frac{\sin^2{\theta}}{\Sigma}[(r^2+a^2)^2-a^2\Delta\sin^2{\theta}]d\phi^{2},
\end{align}   
$\text{where}\;\;\Sigma=r^2+a^2\cos^2{\theta}$,
    $2f=2Mr-hr\ln\left({r}/{2M}\right)$, $\Delta=r^2+a^2-2Mr+h r\ln\left({r}/{2M}\right)$,
and $a$ is the spin parameter. The rotating metric (\ref{metric2}) is governed by three parameters $M$, $a$ and $h$, which measures potential deviation from the Kerr black hole. The metric (\ref{metric2})
encompasses the Kerr black hole in the absence of scalar field ($h=0$) and it will henceforth be referred as rotating Horndeski black hole. Hairy black holes have been well studied in literature viz., the stationary black hole solution with new global charges that are not associated with the Gauss law \citep{Herdeiro:2015waa}, e.g., black holes with proca hair \citep{Herdeiro:2016tmi} or scalar hair \citep{Herdeiro:2014goa,Gao:2021luq}. A recent review of black holes with hair due to global charge can be found in Herdeiro \& Radu (\citeyear{Herdeiro:2015waa}).
The metric (\ref{metric2}) is a prototype non-Kerr black hole that mathematically resembles the 
Kerr metric, with mass $M$ replaced by the mass function \citep{Bambi:2014nta} $$m(r)=M-\frac{h}{2}\ln\left(\frac{r}{2M}\right).$$
Further, the metric (\ref{metric2}), like the Kerr black hole, possesses time translational and rotational invariance isometries that correspond to the existence of Killing vectors 
$\chi_{(t)}^{\mu}=\delta _t^{\mu }$  and $\chi_{(\phi)}^{\mu}=\delta _{\phi }^{\mu }$ respectively.  
The rotating Horndeski black holes are singular at $\Sigma=0$ which corresponds to a ring singularity, whereas the null surface $\Sigma\neq0$ and $\Delta(r)=0$ is a coordinate singularity corresponding to horizon radii which are zeroes of $g^{rr}=0=\Delta(r)$.
For given $a$ and $h$ in the parameter space (see Figure~\ref{parameterSpace}), $\Delta(r)=0$ admits two possible roots--- the Cauchy horizon ($r_{-}$) and the event horizon ($r_{+}$)---which in the limit $h\to 0$ revert to the horizons of the Kerr metric, $r_{\mp}^{Kerr} = M\mp\sqrt{M^2-a^2}$. 

The parameter space ($a, h$) for rotating Horndeski blacks hole is depicted in Figure~\ref{parameterSpace}. The black holes exist when $a<a_E$ ($h>h_E$), as depicted by the shaded region in Figure~\ref{parameterSpace}, and they become Kerr black holes when $a_E=M$ ($h=0$). The points ($a_E, h_E$) on the solid boundary line give extremal rotating Horndeski black holes whereas, for $a>a_E$ ($h<h_E$) one has a naked singularity.  The left panel in Figure~\ref{horizonsFig} depicts the horizon structure of the rotating Horndeski black holes, wherefrom the horizon radii ($r_{\mp})$ are obtained at $\Delta(r)=0$. The right panel of Figure~\ref{horizonsFig} shows constant contours of $\Delta r_{EH}=r_+ -r_{+}^{Kerr}$; $\Delta r_{EH}<0$ in the ($a$-$h$) space elucidates that, for a given spin $a$, the rotating Horndeski black holes have a smaller event horizon radius than the Kerr black holes. For a given spin $a$, there exists an extremal value of $h$, $h_E$  such that $\Delta(r)=0$ has a double root which corresponds to an extremal black hole with degenerate horizons. When $h > h_E$ , $\Delta(r)=0$  has two simple zeros, and it has no zeros for $h < h_E$ (see the left panel of Figure~\ref{horizonsFig}), resulting in , respectively, a non-extremal black hole with a Cauchy horizon and an event horizon, on the one hand, and a no-horizon spacetime, on the other.

The frame dragging effect in the vicinity of the rotating Horndeski black holes (\ref{metric2}) is caused by its non-zero off diagonal elements, i.e., $g_{t\phi}$. Due to this effect, a stationary observer outside the event horizon, moving with zero angular momentum with respect to an observer at spatial infinity, rotates with the black hole with an angular velocity given by \citep{poisson2004relativist}:
\begin{eqnarray}\label{omega1}
    \tilde{\omega}=\frac{d\phi}{d t}=-\frac{g_{t\phi}}{g_{\phi\phi}}=\frac{2ar[M-\frac{h}{2}\ln{(\frac{r}{2M})}]}{(r^2+a^2)^2-a^2 \Delta\sin^2{\theta}}.
\end{eqnarray}
The velocity $\tilde{\omega}$ increases monotonically as the observer approaches the black hole, and at the event horizon the observer begins maximally co-rotating with a velocity equal to that of the black hole, which is given by
\begin{eqnarray}\label{omega2}
    \Omega= \tilde{\omega} |_{r=r_{+}}=\frac{2ar_{+}[M-\frac{h}{2}\ln{(\frac{r_{+}}{2M}})]}{(r_{+}^2+a^2)^2}, 
\end{eqnarray}
which, in the limit $h\to0$, reduces to the angular velocity of a Kerr black hole. The surface of the black hole is rotating as a rigid body \citep{Frolov:2014dta}, in a sense that each point of the horizon has the same angular velocity (as measured at infinity).

\begin{figure}[t]
\centering
    \includegraphics[scale=0.8]{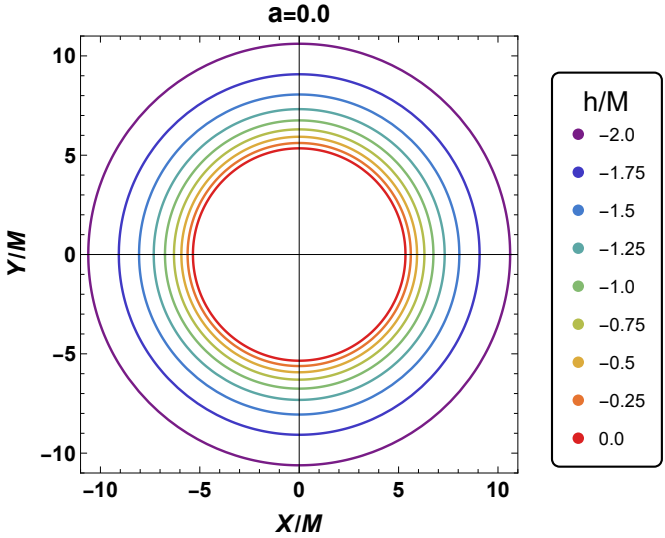}
\caption{Shadows of spherical Horndeski black holes ($a=0$) with  equispaced values of $h$. The shadows are sparsely spaced with increasing $|h|$. }\label{shadow1}
\end{figure}
\begin{figure*}
\centering
\begin{tabular}{c c}
    \hspace{-2cm}\includegraphics[scale=0.85]{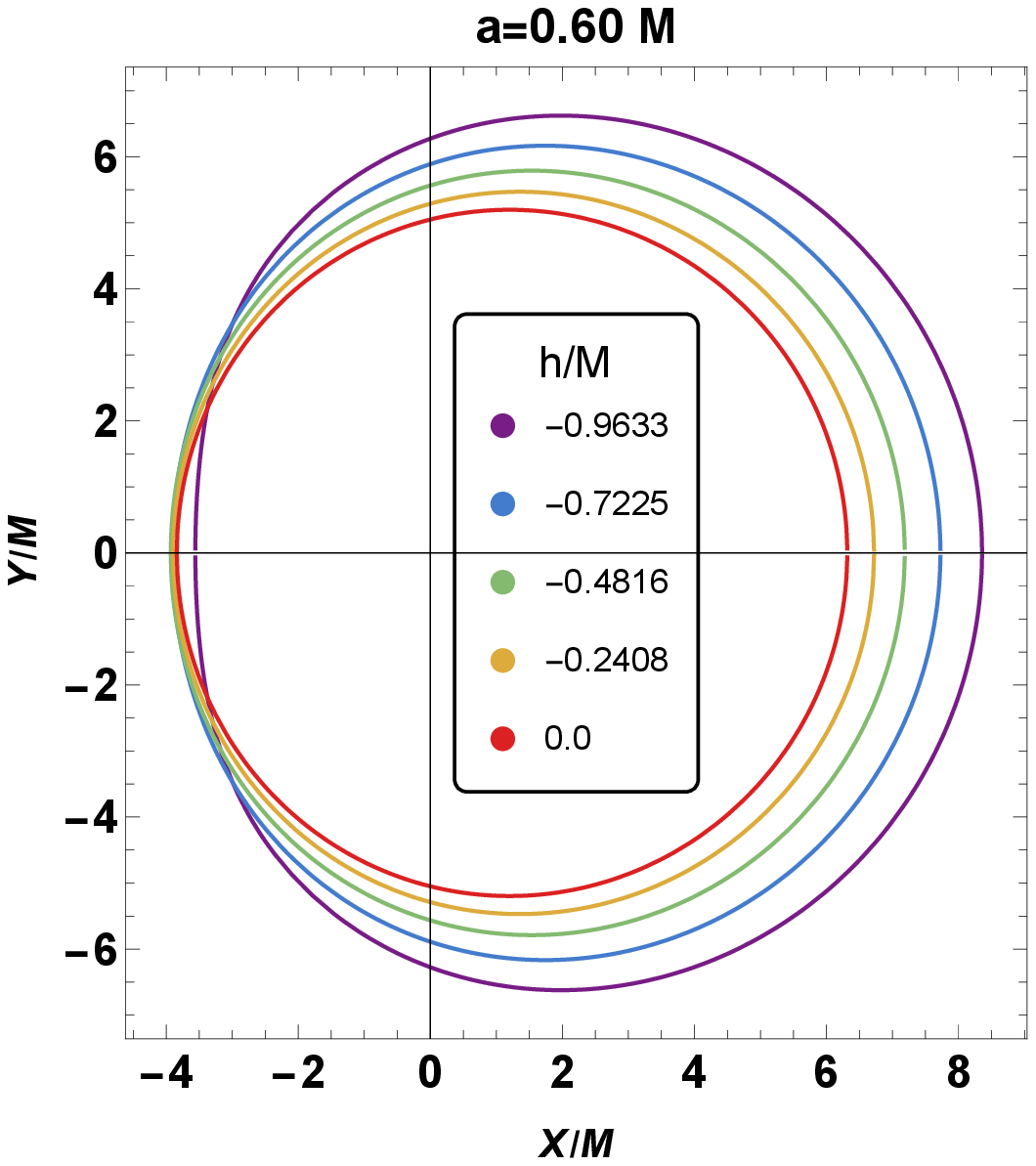}&
    \includegraphics[scale=0.85]{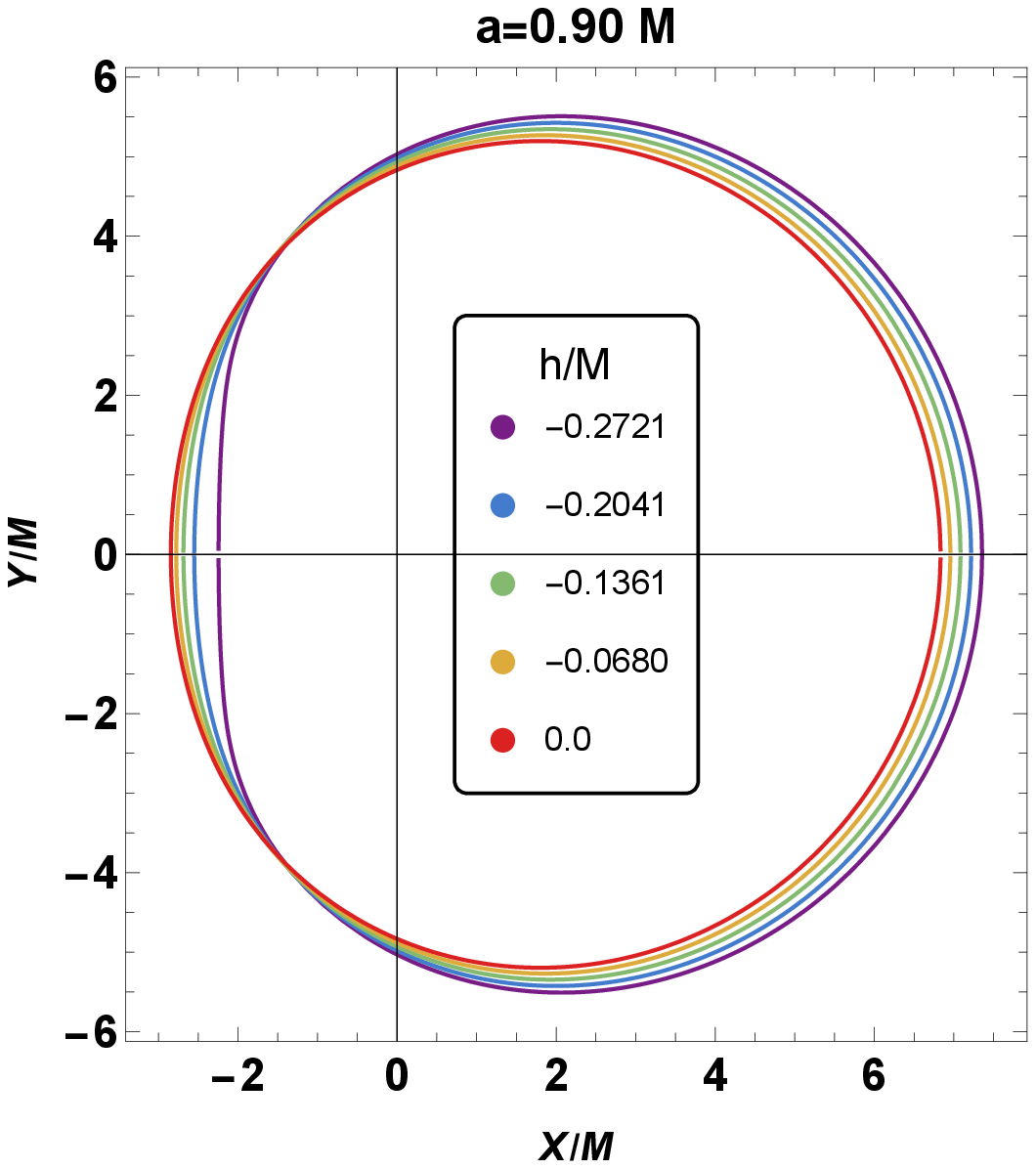}
\end{tabular}
\caption{Shadows of rotating Horndeski black holes. The parameter space (see Figure~\ref{parameterSpace}) constrains the maximum value of $|h|$ for a given $a$, e.g., $h_E=-0.9833 M, -0.2771 M$ for $a= 0.6 M, 0.9 M$ respectively. }\label{shadow2}
\end{figure*}

\section{Spherical photon orbits and black hole shadows}\label{Sec3}
The geodesics around the black hole, while being precursors to the study of phenomena such as strong field gravitational lensing \citep{Kumar:2020sag,Kumar:2020ltt,Islam:2020xmy}, accretion processes, and shadow formation, carry more intrinsic significance in a sense that they bear an imprint of the theory of gravity on which the spacetime is laid \citep{Chandrasekhar:1985kt,Cunha:2018gql, Bambi:2019tjh,Kumar:2020owy,Kumar:2020hgm,Kumar:2019pjp,Kumar:2019ohr,Kumar:2017tdw, Afrin:2021imp,Younsi:2021dxe}. The spherical timelike orbits around Kerr black holes pioneered by Wilkins \citep{Wilkins:1972rs}, the spherical null geodesics around Kerr black holes investigated by Teo \citep{Teo:2020sey}, and the photon region \citep{Johnson:2019ljv} are imperative to the understanding of shadow formation thereon. 
The metric (\ref{metric2}), which is invariant under time translational and rotational invariance, entails two Killing vector fields, $\chi_{(t)}^{\mu}$ and $\chi_{(\phi)}^{\mu}$, and conserved quantities energy $\mathcal{E}$ and angular momentum $\mathcal{L}$ respectively. Carter's approach for solving the Hamilton–Jacobi equation then leads to the first order differential equations of photon motion \citep{Chandrasekhar:1985kt}:
\begin{align}
\Sigma \frac{dt}{d\lambda}=&\frac{r^2+a^2}{\Delta}[\mathcal{E}(r^2+a^2)-a\mathcal{L}]-a(a\mathcal{E}\sin^2{\theta}-\mathcal{L}),\label{TimeEq}\\
\Sigma \frac{dr}{d\lambda}=&\pm\sqrt{\Re(r)}\ ,\label{RadialEq} \\
\Sigma \frac{d\theta}{d\lambda}=&\pm\sqrt{\Theta(\theta)}\ ,\label{Theta}\\
\Sigma \frac{d\phi}{d\lambda}=&\frac{a}{\Delta}[\mathcal{E}(r^2+a^2)-a\mathcal{L}]-\left(a\mathcal{E}-\frac{\mathcal{L}}{\sin^2{\theta}}\right)\label{PhiEq},
\end{align}
where $\lambda$ is the affine parameter. The radial and polar potential functions ${\Re}$ and $\Theta$ are respectively given by\citep{Chandrasekhar:1985kt}
\begin{eqnarray}
 {\Re}&=&[(r^2+a^2)\mathcal{E}-a\mathcal{L}]^2-\Delta [{\mathcal{K}}+(a \mathcal{E}-\mathcal{L})^2],\label{radialPotential}\\ 
 \Theta&=&\mathcal{K}-\left(\frac{{\mathcal{L}}^2}{\sin^2\theta}-a^2 \mathcal{E}^2 \right)\cos^2\theta\ .\label{thetaPotential}
\end{eqnarray}
Here, $\mathcal{K}$ is the separability constant related to the Carter constant $\mathcal{Q}$---associated with a non-apparent spacetime symmetry---through $\mathcal{K}=\mathcal{Q}-(a\mathcal{E}-\mathcal{L})^2$ \citep{Carter:1968rr,Chandrasekhar:1985kt}. Moreover, we introduce two energy rescaled parameters, $ \xi=\mathcal{L}/\mathcal{E}$ and $\eta=\mathcal{K}/\mathcal{E}^{2}$, the \textit{critical impact parameters}.
For a spherical photon orbit (SPO) at radius $r_{p}$, the photon must have a radial turning point given by $\Dot{r}=0$, $\Ddot{r}=0$, which further imply, ${\Re}=0$ and ${\Re}'=0$ \citep{Teo:2020sey}, and by using Equation~(\ref{radialPotential}) we obtain the critical impact parameters, 
\begin{widetext}
\begin{eqnarray}
           \xi_{c} & =& - \frac{h (3 r^2-a^2) \ln \left(\frac{M r}{2}\right)-a^2 (h-2 M-2 r)-r^2 (h
           + 6 M-2 r)}{a [h \ln \left(\frac{M r}{2}\right)+h-2 M+2 r] \label{xiCritical}} \\
           \eta_{c}=&-&\frac{r^3 \Big[2 h \ln \left(\frac{M r}{2}\right) \left(4 a^2-3 r (h+6 M-2 r)\right)
         -8 a^2 (h+2 M)+9 h^2 r \ln ^2\left(\frac{M r}{2}\right)+r (h+6 M
         -2 r)^2\Big]}{ a^2 [h \ln \left(\frac{M r}{2}\right)+h-2 M+2 r]^2}.\label{etaCritical}
\end{eqnarray}
\end{widetext}
Equations~(\ref{xiCritical}) and (\ref{etaCritical}), when $h\to 0$, reduce to those of the Kerr case ($ \xi_{c}^k$, $\eta_{c}^k$) \citep{Chandrasekhar:1985kt}. The photon shell is the region of a black hole spacetime containing bound null geodesics---while it is a 2-sphere with radius $3M$ for Schwarzschild black holes, for Kerr black holes it becomes a spherical shell, such that \citep{Teo:2020sey,Johnson:2019ljv}
\begin{equation}
	r_{p}^{-}\leq r_p\leq r_{p}^{+};\;\;
    r_{p}^{\mp}\equiv2M\left[1+\cos\left({\frac{2}{3}\arccos\left(\mp\frac{|a|}{M}\right)}\right)\right]\,
\end{equation}
where $r_p^\mp$ are, respectively, the prograde and the retrograde photon radii: $\eta_{c}^k=0$, $\xi_{c}^k(r_p^\mp)\gtrless0$. While a unique bound orbit passes through every point in the equatorial annulus---$r_{p}^{-}\leq r\leq r_{p}^{+}$, $\theta=\pi/2$---the orbits are planar and confined to the equatorial plane on the boundaries $r=r_p^\mp$ \citep{Teo:2020sey}.
The $\xi_{c}^k$ is related to the angular momentum of the photon about the $\phi$-axis\citep{Teo:2020sey}; for photons with zero angular momenta, the overall direction of the orbits reverses at the intermediate value $r_p^0$ which can be determined by zeros of $\xi_{c}^k=0$. At generic points, the spherical photons oscillate in the $\theta$-direction between polar angles \citep{Johnson:2019ljv}:
\begin{equation}\label{Turning_Points}
	\theta_\pm=\arccos\left(\mp\sqrt{\nu_+}\right),
\end{equation}
where
\begin{align}
	\nu_\pm=&\frac{r}{a^2(r-M)^2}\big[-r^3+3M^2r-2a^2M \nonumber\\
	\pm&2\sqrt{M(r^2+a^2-2Mr)(2r^3-3Mr^2+a^2M)}\big].
\end{align}
At the radius $r=r_p^0$, $\{\theta_-,\theta_+\}=\{0,\pi\}$ and the orbits can cross the poles.
Thus, the photon shell can be summarized as set of all spacetime points:
$r_{p}^{-}\leq r\leq r_{p}^{+},\;
	\theta_-\le\theta\le\theta_+,\; 
	0\le\phi<2\pi,\;
	-\infty\le t\le\infty.$
The bound SPOs are unstable, i.e., at $r=r_p\in(r_p^-, r_p^+)$, the ${\Re}''\leq0$
and a slight perturbation will result in an exponential divergence of the photon away from its spherical orbit. Orbits with slightly smaller $r$ plunge into the black hole, while slightly larger orbits escape to infinity. The observed photon ring image arises from photons traveling on such nearly bound geodesics \citep{Johnson:2019ljv}. 

The photon rings in the limit of vanishing thickness \citep{Johnson:2019ljv}, reduce to the projection along the SPOs---whose inner edges, marked by a sharp flux fall-offs, outline the geometrical shadow of the black hole \citep{Johannsen:2013vgc}. The black hole shadow is the projection of photon sphere as observed at spatial infinity. The shadow shape depends on the black hole parameters, i.e., spin and other hairs \citep{Johannsen:2015mdd,Afrin:2021imp,Afrin:2021ggx} alongside the observation angle $\theta_{0}$ relative to the spin axis, with the overall size scaled by the black hole mass $M$ \citep{Akiyama:2019bqs}. Thus, at radial infinity and an inclination angle $\theta_0$, an observer can visualize the black hole shadow outlined by the celestial coordinates defined by \citep{Bardeen:1973tla,Frolov:1418196,Kumar:2018ple}
\begin{equation}
\{X,Y\}=\{-\xi_{c}\csc\theta_o,\, \pm\sqrt{\eta_{c}+a^2\cos^2\theta_o-\xi_{c}^2\cot^2\theta_o}\}\,\label{Celestial1}
\end{equation}
The effect of spin is more significant for an equatorial observer ($\theta_o=\pi/2$), in which case Equation~(\ref{Celestial1}) simplifies to
\begin{eqnarray}
\{X(r_p),Y(r_p)\}=\{-\xi_{c}(r_p),\, \pm\sqrt{\eta_{c}(r_p)}\},\label{Celestial2}
\end{eqnarray}
which satisfies $X^2+Y^2=\xi_{c}^2 + \eta_{c}$. The black hole shadow can be constructed by plotting ($X$, $Y$). The spherical Horndeski black hole shadows, depicted in Figure~\ref{shadow1}, have slightly larger radii than the Schwarzschild shadow radius $3\sqrt{3}M$. It turns out that the rotating Horndeski black hole shadows are significantly different from the Kerr shadows (see Figure~\ref{shadow2}), and the parameter $h$ has a profound influence on them. For any spin parameter $a$, the shadow size becomes larger and more distorted with increasing $|h|$. We also notice a horizontal shift in shadow along the $x$-axis, with increase in $|h|$ and $a$, due to the frame dragging effect. 
Interestingly, the influence of the $h$ parameter on the shadow deformation is similar to that caused by the spin $a$ on the Kerr black hole shadow; hence, it is likely that the rotating Horndeski black holes for some parameters ($a$, $h$) may mimic the Kerr black holes. 
\section{Black hole Parameters estimation}\label{Sec4}
 While the first order correction to the circular shadow shape occurs due to the spin in Kerr spacetime, in MoGs the distortion for a given spin may arise due to other hairs \citep{Cunha:2015yba,Cunha:2019ikd,Ghosh:2020spb,Afrin:2021imp,Khodadi:2021gbc}---this prompts the use of shadow observables for determination of black hole parameters \citep{Hioki:2009na,Kumar:2018ple,Afrin:2021imp,Afrin:2021ggx}. It has been shown in multiple studies that observables like shadow radius $R_s$ and distortion $\delta_s$ \citep{Hioki:2009na} demand some specific symmetry in the shadow shape, and thus may not be efficient in some MoGs \citep{Abdujabbarov:2015xqa,Tsukamoto:2014tja,Kumar:2018ple}. Kumar and Ghosh \citep{Kumar:2018ple} proposed estimating the black hole parameters from haphazard shadow shapes---the prescription was subsequently used to estimate the parameters associated with several rotating black holes in MoGs \citep{Kumar:2017tdw,Afrin:2021imp}.

\begin{figure}[t]
            \centering
			\includegraphics[scale=0.9]{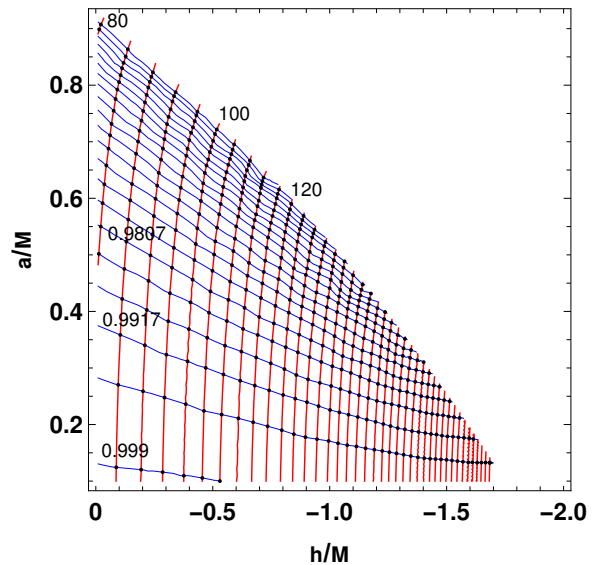}
	\caption{Contour plots of the observables $A/M^2$ (red curves) and $D$ (blue curves) in the parameter space ($a/M,h/M$). $A/M^2$: 80 to 240 in steps of 4; $D$: 0.926 to 0.999 in steps of 0.00365.}
	\label{parameterestimation}
\end{figure}
\begin{table}[t]
    \caption{Estimated values of the Parameters $a/M$ and $h/M$ for the Rotating Horndeski black holes}
      \centering
       \begin{tabular}{c c c c} 

 $A/M^2$ & $D$ & $a/M$ & $h/M$ \\ [0.8ex] 
 \hline\hline
 \;\; 84\;\; &\;\; 0.94060\;\; &\;\; 0.8119\;\; &\;\; -0.1090\;\; \\ [1ex] 
 \;\; 92\;\; &\;\; 0.95520\;\; &\;\; 0.6816\;\; &\;\; -0.2899\;\; \\ [1ex] 
 \;\; 100\;\; &\;\; 0.96980\;\; &\;\; 0.5484\;\; &\;\; -0.4424\;\; \\ [1ex] 
 \;\; 120\;\; &\;\; 0.98805\;\; &\;\; 0.3071\;\; &\;\; -0.7545\;\; \\ [1ex] 
 \;\; 192\;\; &\;\; 0.99535\;\; &\;\; 0.1383\;\; &\;\; -1.4350\;\; \\ [1ex] 
 \hline
\end{tabular}
\label{parameterestimation_table}
\end{table}

The area enclosed by the black hole shadow is \citep{Kumar:2018ple}
\begin{eqnarray}
A&=&2\int{Y(r_p) dX(r_p)}\nonumber\\
&=&2\int_{r_p^{-}}^{r_p^+}\left( Y(r_p) \frac{dX(r_p)}{dr_p}\right)dr_p,\label{Area}
\end{eqnarray} 
where the prefactor 2 is due to the symmetry about the $x-$axis,
whereas, the oblateness ($D$) can be written as \citep{Kumar:2018ple},
\begin{eqnarray}
D=\frac{X_r-X_l}{Y_t-Y_b}\label{Oblateness}
\end{eqnarray}
where the subscripts $l$, $r$, $t$ and $b$ stand for the left and right ends of the shadow silhouette, where $Y(r_p)=0$ (considering positive $a$), and the top and bottom points, where $Y'(r_p)=0$ respectively \citep{Hioki:2009na}. While for a spherically symmetric black hole $D=1$, however,  $\sqrt{3}/2\leq D<1$ for the Kerr black hole \citep{Tsupko:2017rdo}.
 \begin{figure*}[t]
		\begin{tabular}{c c}
			\hspace{-1cm}\includegraphics[scale=0.6]{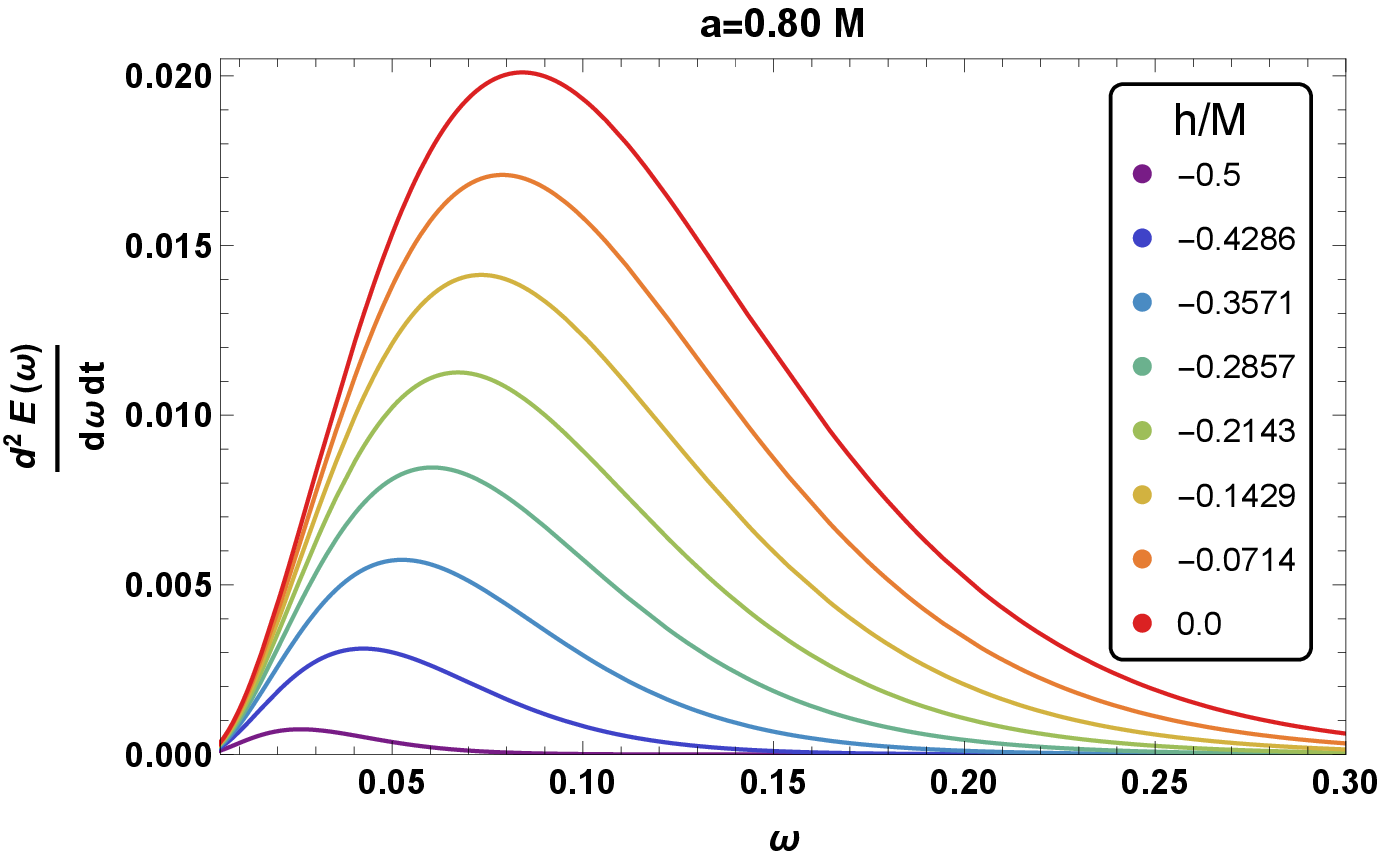}&
			\includegraphics[scale=0.6]{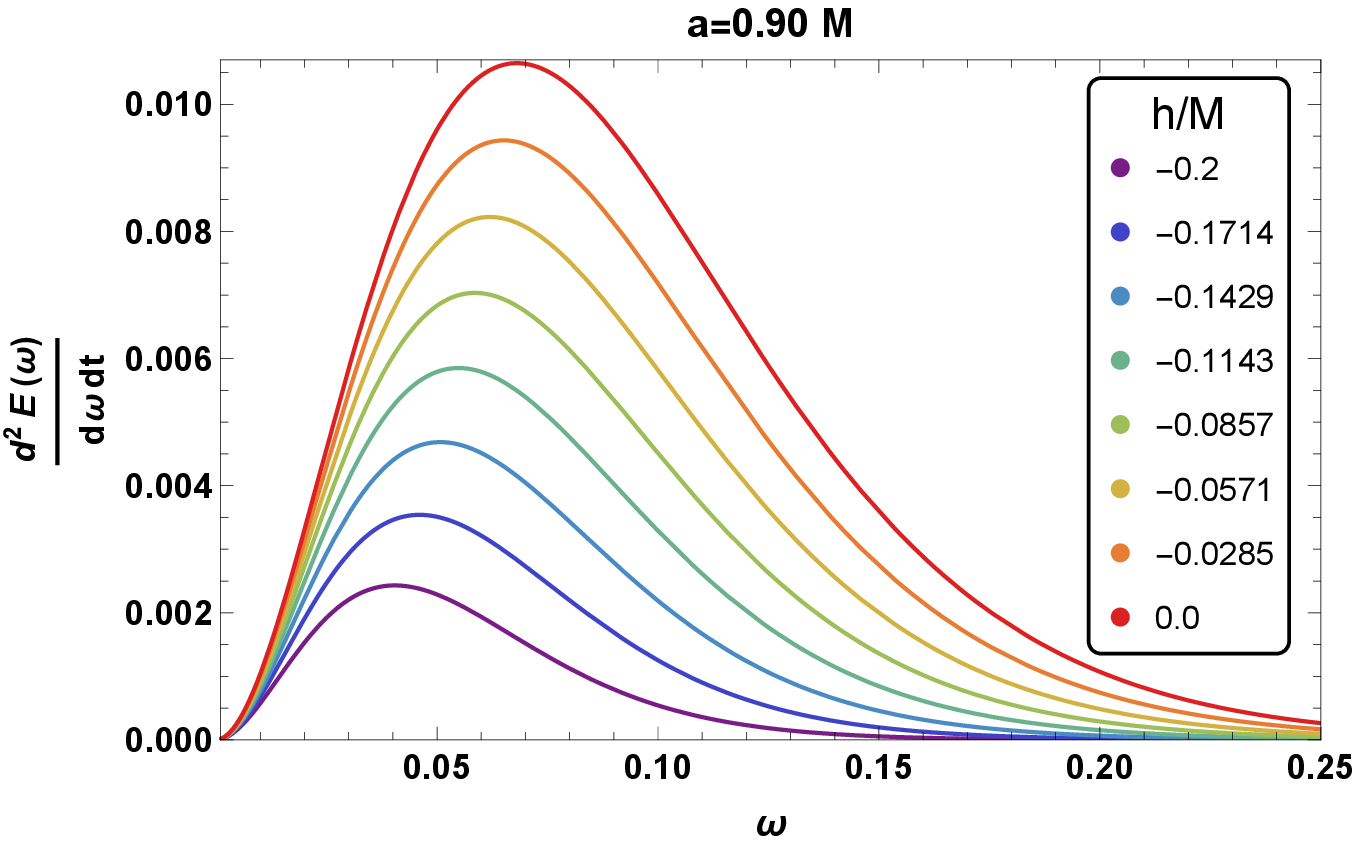}
		\end{tabular}	
	\caption{Evolution of the emission rate with frequency $\omega$ for different values of parameter $h$.}
	\label{EnergyEmission}
\end{figure*}
 
Evidently from the shadow structure (see Figure~\ref{shadow2}), both the parameters $a$  and $h$ have profound impact on the shadow area as well as on the oblateness. There maybe the possible degeneracy of the $A$ and $D$ in the $a$ and $h$, in the sense that two or more combinations of the black hole parameters ($a$, $h$) may give the same $A$ \textit{and/}\textit{or} $D$ as scrutinized in Figure~\ref{parameterestimation}, wherefrom it is found that, (i) a contour of a given observable (A or D) gives one-to-one correspondence between $a$ and $h$ parameters, and (ii) the contours of the two different observables (A and D) intersect at unique points.
This leads to the inference that both the shadow observables $A$ and $D$ are non-degenerate in parameters ($a, h$) if at least one of the two parameters are fixed. The observables are also degenerate for an infinite number of unique parameter points lying on a given constant contour curve. From the observation \textit{(ii)} we surmise that from each intersection point (the black points in Figure~\ref{parameterestimation}) of the $A$ and $D$ contour lines, one can uniquely determine the parameters $a$ and $h$ of the black holes that we tabulate in Table \ref{parameterestimation_table}.
\subsection{Energy emission}
We focus on the energetic aspects by examining the energy emission rate.
It has been shown that the absorption cross-section approaches the black hole shadow for a distant observer, which oscillates around a constant limiting value $\sigma_{lim}$, which is same as the geometrical cross-section of the photon sphere of the black hole  \citep{Wei:2013kza,Amir:2016cen,Belhaj:2020okh}, as
$\sigma_{lim}\approx\pi R_{s}^2,$
where $R_s$ designates the radius of the shadow approximated by a reference circle and is given by \citep{Hioki:2009na}:
\begin{eqnarray}
    R_s=\frac{(X_t-X_r)^2+Y_{t}^2}{2|X_r-X_t|},
\end{eqnarray}\label{Rs}
using the relations $X_b=X_t$ and $Y_b=-Y_t$ \citep{Hioki:2009na}.
The energy emission rate of a rotating black hole is given by Wei \& Liu \citeyear({Wei:2013kza}), Amir \& Ghosh \citeyear({Amir:2016cen}), and Belhaj et al. \citeyear({Belhaj:2020okh}):
\begin{eqnarray}\label{emissionEq} 
\frac{d^2E(\omega)}{d\omega dt}=\frac{2\pi^2R_{s}^2}{e^{\omega/T_+}-1}\omega^3,
\end{eqnarray}
where $\omega$ is photon frequency and $T_+$ is the Hawking temperature at event horizon $r_{+}$ given by
\begin{equation}
    T_+=\displaystyle{\lim_{r\to r_+}}\frac{1}{2\pi} \frac{\partial_r \sqrt{g_{tt}}}{\sqrt{g_{rr}}}.
\end{equation}
For the rotating Horndeski black holes, we find
\begin{align}
    T_+=&\frac{1}{4\pi \left(a^2+r_{+}^2\right)^2} \Big[h \left(a^2-r_{+}^2\right) \ln \left(\frac{M r_{+}}{2}\right) \nonumber \\
    +& a^2 (h-2 M)+r_{+}^2 (h+2 M)\Big]. \label{hawkingTempEq}
\end{align}
The energy emission rate decreases with increasing $|h|$, while the Gaussian peak shifts to lower $\omega$. Moreover, the spin parameter $a$ decreases the energy emission rate (see Figure~\ref{EnergyEmission}). Interestingly, the Hawking temperature and the energy emission rate, in the limit $h \to 0$, go over to those of the Kerr black hole. 
\section{Constraints from EHT}\label{Sec5}
\begin{figure*}
\begin{center}
    \begin{tabular}{c c}
    \includegraphics[scale=0.80]{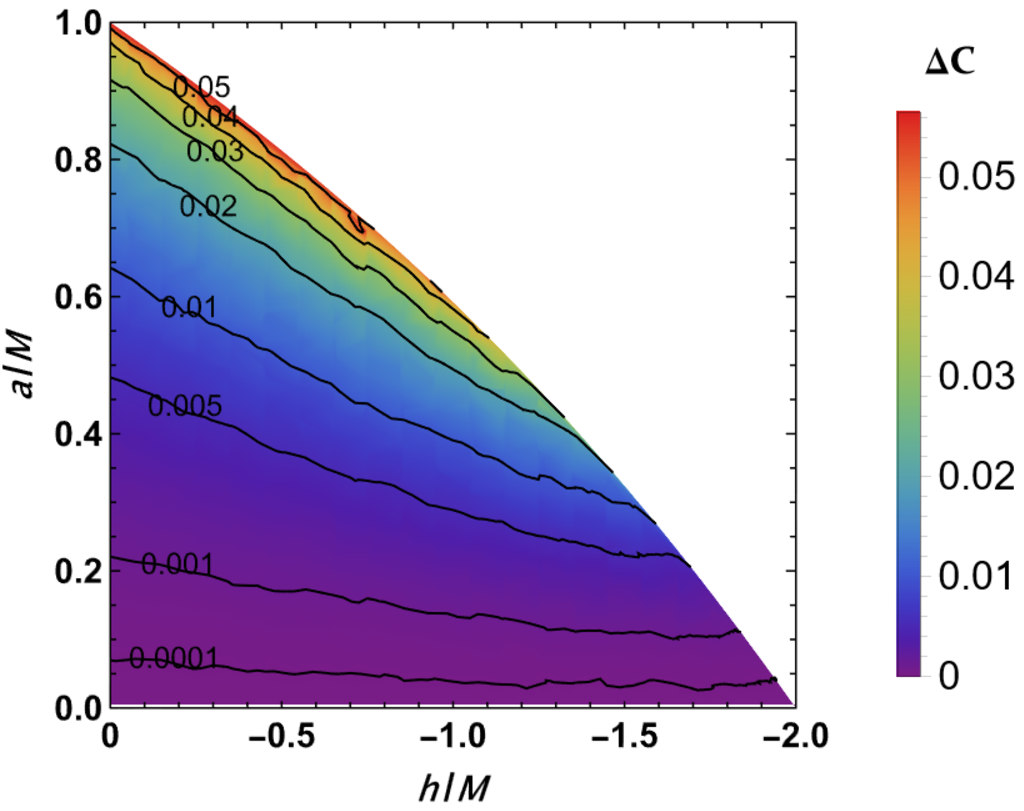}&
     \includegraphics[scale=0.80]{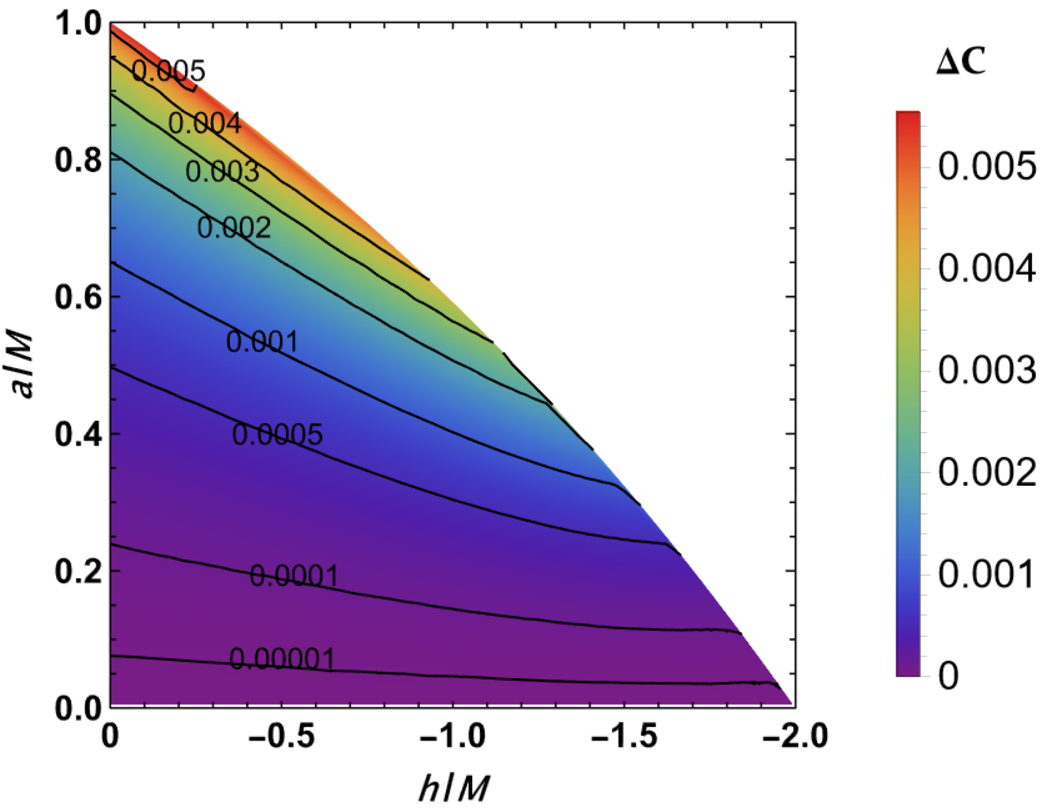}
\end{tabular}
\end{center}
	\caption{The circularity deviation observable $\Delta C$ for the rotating Horndeski black hole shadows as a function of parameters ($a/M$ and $h/M$), in agreement with the EHT observations of the M87* black hole, i.e., $\Delta C \leq 0.1$ is satisfied for the entire parameter space ($a/M$ and $h/M$). The mass and distance of M87* used are $M=6.5\times 10^9 M_{\odot}$ and $d=16.8$ Mpc. The inclination angle is $\theta_0=90$\textdegree (left) and $\theta_0=17$\textdegree (right). The white region is forbidden for ($a/M$ and $h/M$).}
	\label{M87obs1}
\end{figure*}
\begin{figure*}
\begin{center}
    \begin{tabular}{c c}
    \includegraphics[scale=0.80]{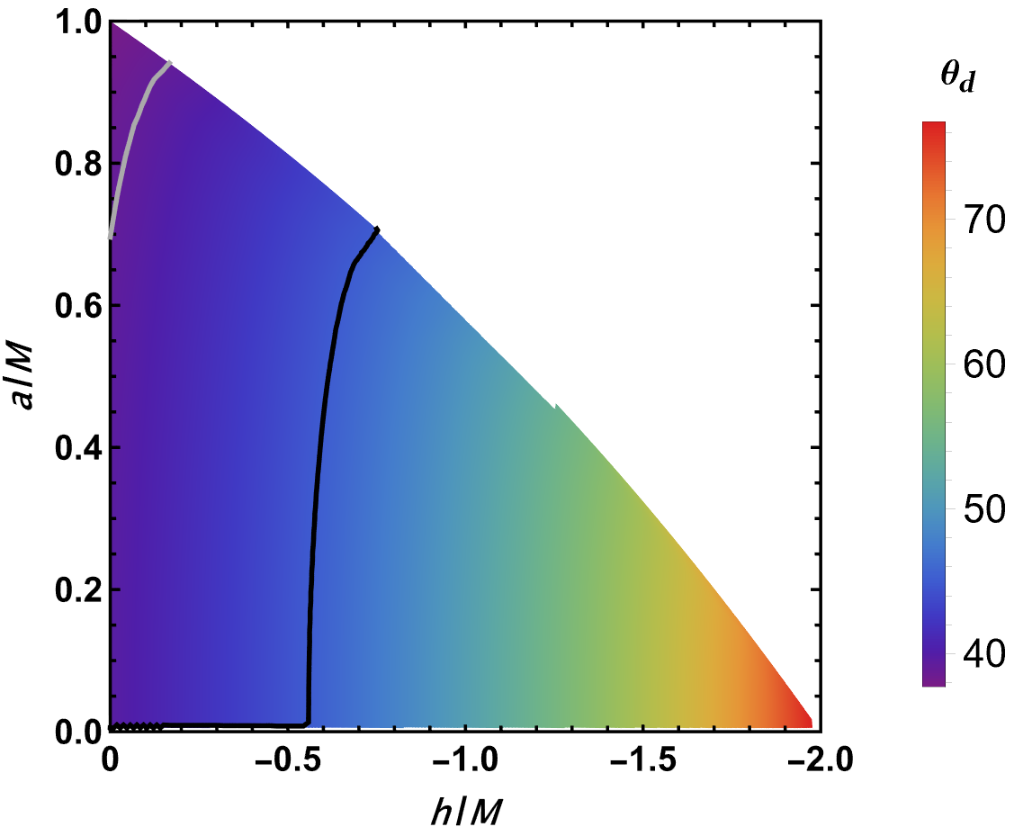}&
     \includegraphics[scale=0.80]{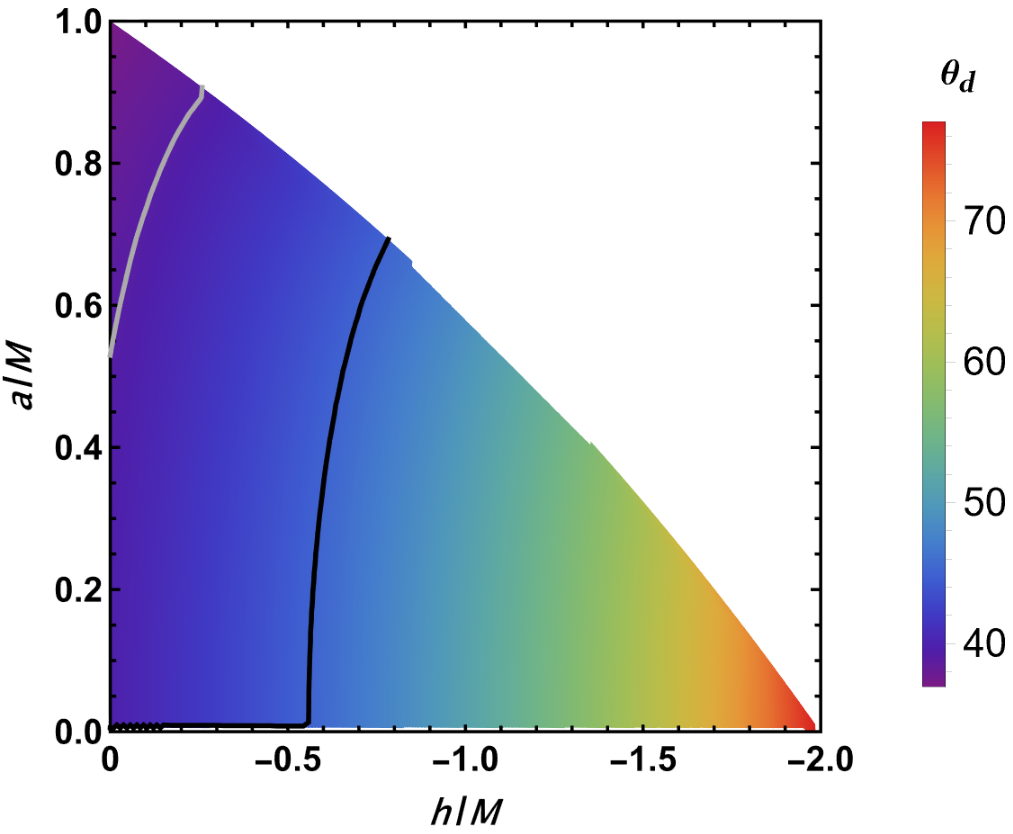}
\end{tabular}
\end{center}
	\caption{Angular diameter observable $\theta_d$ for  rotating Horndeski black hole shadows as a function of parameters ($a/M$ and $h/M$). The light gray and black solid curves correspond to $39\mu$as and $45\mu$as, respectively. The mass and distance of M87* used are $M=6.5\times 10^9 M_{\odot}$ and $d=16.8$Mpc. The inclination angle is $\theta_0=90$\textdegree (left) and $\theta_0=17$\textdegree (right). The white region is forbidden for ($a/M$ and $h/M$).}
	\label{M87obs2}
\end{figure*}
We next investigate the possible constraints on Horndeski gravity using the shadow of black hole M87* via EHT observations.
The EHT collaboration  \citep{Akiyama:2019cqa,Akiyama:2019bqs,Akiyama:2019eap} has released the first image of supermassive black hole M87*; as such, it may be possible to probe deeper into the strong-field regime of gravity and eventually test the no-hair theorem \citep{Carter:1971zc}. The observed image shows a ring of diameter  $\theta_d=42\pm3\mu$as and  deviation from circularity $\Delta C\lesssim 0.1$,  as per the shadow of a Kerr black hole, which, according to the Kerr hypothesis \citep{Psaltis:2007cw}, describes the background spacetime of an astrophysical black hole. The Kerr hypothesis, a strong-field prediction of GR, may be violated in the MoGs that can also admit non-Kerr black holes \citep{Berti:2015itd}. It has been shown in previous studies that the Kerr metric remains a solution in some alternative theories of gravity \citep{Psaltis:2007cw}. Alongside $\theta_d$, the measured circularity deviation  $\Delta C$ for the M87* black hole shadow can also constrain the black hole parameters \citep{Kumar:2020hgm,Bambi:2019tjh}.
Thus, presupposing the M87* a rotating Horndeski black hole, and using the EHT constraints on two shadow observables, the deviation from circularity $\Delta C$ and the angular diameter $\theta_d$ of the black hole shadow, we investigate the constraints for the rotating Horndeski black hole to be a suitable candidate for the M87* black hole. We shall take the mass of M87* as reported by the EHT collaboration, $M=6.5\times 10^9 M_{\odot}$ and the distance $d=16.8$ Mpc \citep{Akiyama:2019cqa,Akiyama:2019bqs,Akiyama:2019eap}.

We first construct a shadow observable: the deviation from  circularity $\Delta C$. The boundary of the black hole shadow is outlined by the polar coordinates ($R(\varphi), \varphi$), with the shadow's centre ($X_c$, $Y_c$) at $X_c=(X_r-X_l)/2$ and $Y_c=0$; thus the shadow admits reflection symmetry about the $x$-axis. The average shadow radius $\bar{R}$ can be written as \citep{Bambi:2019tjh} 
\begin{equation}
\bar{R}^2=\frac{1}{2\pi}\int_{0}^{2\pi} R^2(\varphi) d\varphi,
\end{equation}
where $R(\varphi)=\sqrt{(X-X_{c})^2+(Y-Y_{c})^2}$ is the radial distance from the shadow centre ($X_c$, $Y_c$) with any point ($X$, $Y$) on the boundary and 
$\varphi\equiv \tan^{-1}[{Y}/({X-X_C})]$ is the subtended polar angle.
We define the circularity deviation $\Delta C$ in terms of root-mean-square distance from an average radius as \citep{Bambi:2019tjh, Afrin:2021imp}
\begin{eqnarray}\label{circularity}
\Delta C=\frac{1}{\bar{R}}\sqrt{\frac{1}{2\pi}\int_0^{2\pi}\left(R(\varphi)-\bar{R}\right)^2d\varphi}.
\end{eqnarray}
The celestial coordinates depend on the parameters associated with the black holes, i.e., mass $M$, spin parameter $a$, parameter $h$ and inclination angle $\theta_o$; hence, the $\bar{R}$ and $\Delta C$ also depend upon these parameters. We use the EHT observational result ($\Delta C\lesssim 0.1$) to place constraints on the parameter space of the rotating Horndeski black holes; i.e., using the definition (\ref{circularity}), we make a comparison between our theoretical prediction and the EHT observation to discern the observationally favoured values of the black hole parameters. Also, taking into consideration the orientation of the magnetohydrodynamic relativistic jets in M87* image, the inclination angle with respect to observational line of sight is estimated to be $163$\textdegree \citep{Walker:2018muw}, but the shadow is maximally deformed only at very high inclination, viz., $\theta_0\approx90$\textdegree. Since the present analysis does not consider the accretion flow, as only the analytic shadow curve is utilized, on account of the top-bottom symmetry of the shadow, the $163$\textdegree~ inclination is equivalent to $17$\textdegree~.
The circularity deviation is depicted in Figure~\ref{M87obs1}, and it is clearly influenced by both the $h$ and the $a$ parameters; also, it increases with the the inclination angle (see Figure~\ref{M87obs1}). Here, we adopt the M87* as a rotating Horndeski black hole and demonstrate that for appropriate $h$ and $a$ parameters, it is possible to produce the shadow of M87*. However, the obtained circularity deviation $\Delta C<0.06$ is much smaller in the allowed parameter space ($a$, $h$) of the rotating Horndeski black holes (see Figure~\ref{M87obs1}).

Next, the angular diameter of the shadow \citep{Kumar:2020owy,Kumar:2017tdw} is given by
\begin{eqnarray}
\theta_d=2\frac{R_a}{d}\;,\;R_a=\sqrt{A/\pi},\label{angularDiameterEq}
\end{eqnarray}  
where $d$ is the distance from M87* to earth and we take $d=16.8$Mpc. In Figure~\ref{M87obs2} we demonstrate that $39\mu as\leq\theta_d \leq 45\mu as$ when $0.0077 M\leq a\leq 0.9353 M$, $-0.7564 M\lesssim h<0$ at $\theta_o=90$\textdegree and $0.0048 M\leq a\leq 0.9090 M$, $-0.7920 M\lesssim h<0$ at $\theta_o=17$\textdegree, where the rotating Horndeski black holes shadows are consistent with the shadow of M87*. Thus, in this constrained parameter space, the M87* can be a rotating Horndeski black hole.

Hence, the consistency of the rotating Horndeski black holes with the M87* observations at infinite possible parameter points ($a$, $h$) within the constrained parameter space elucidates the fact that they can be strong candidates for astrophysical black holes and, thereby, that the Horndeski gravity, if it is distinguishable from GR, would put the Kerr hypothesis to an astrophysical test. 
Hereon, in the following section, we will conduct a systematic bias analysis between the shadows of rotating Horndeski black holes and the Kerr black hole within this constrained parameter space, to quantify their distinguishability and explore the possibility of testing GR against the Horndeski theory.
\section{Systematic bias analysis within EHT constrained parameter space}\label{Sec6}
\begin{figure*}
\begin{center}
    \begin{tabular}{c c}
    \includegraphics[scale=0.6]{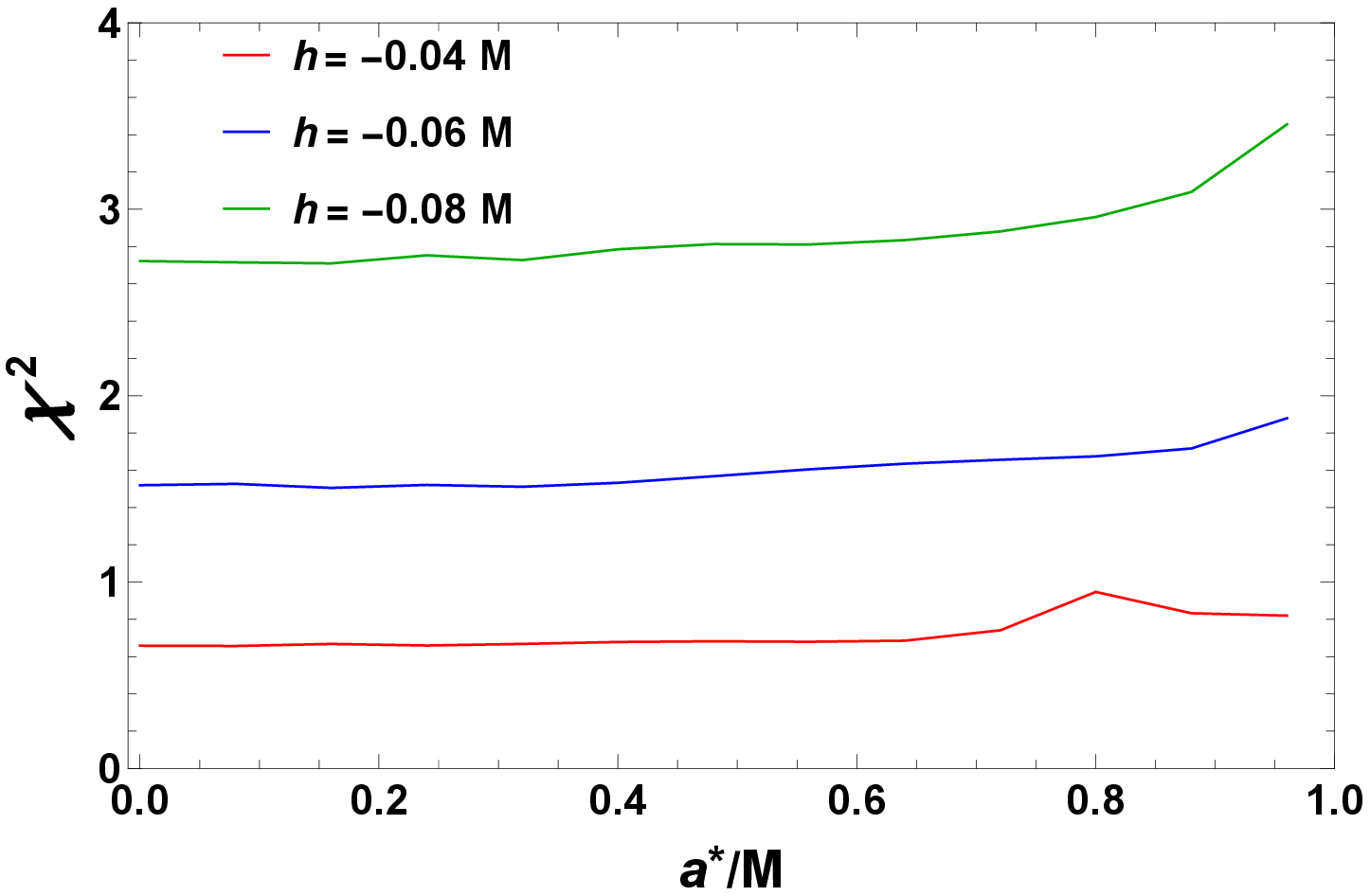}&
     \includegraphics[scale=0.6]{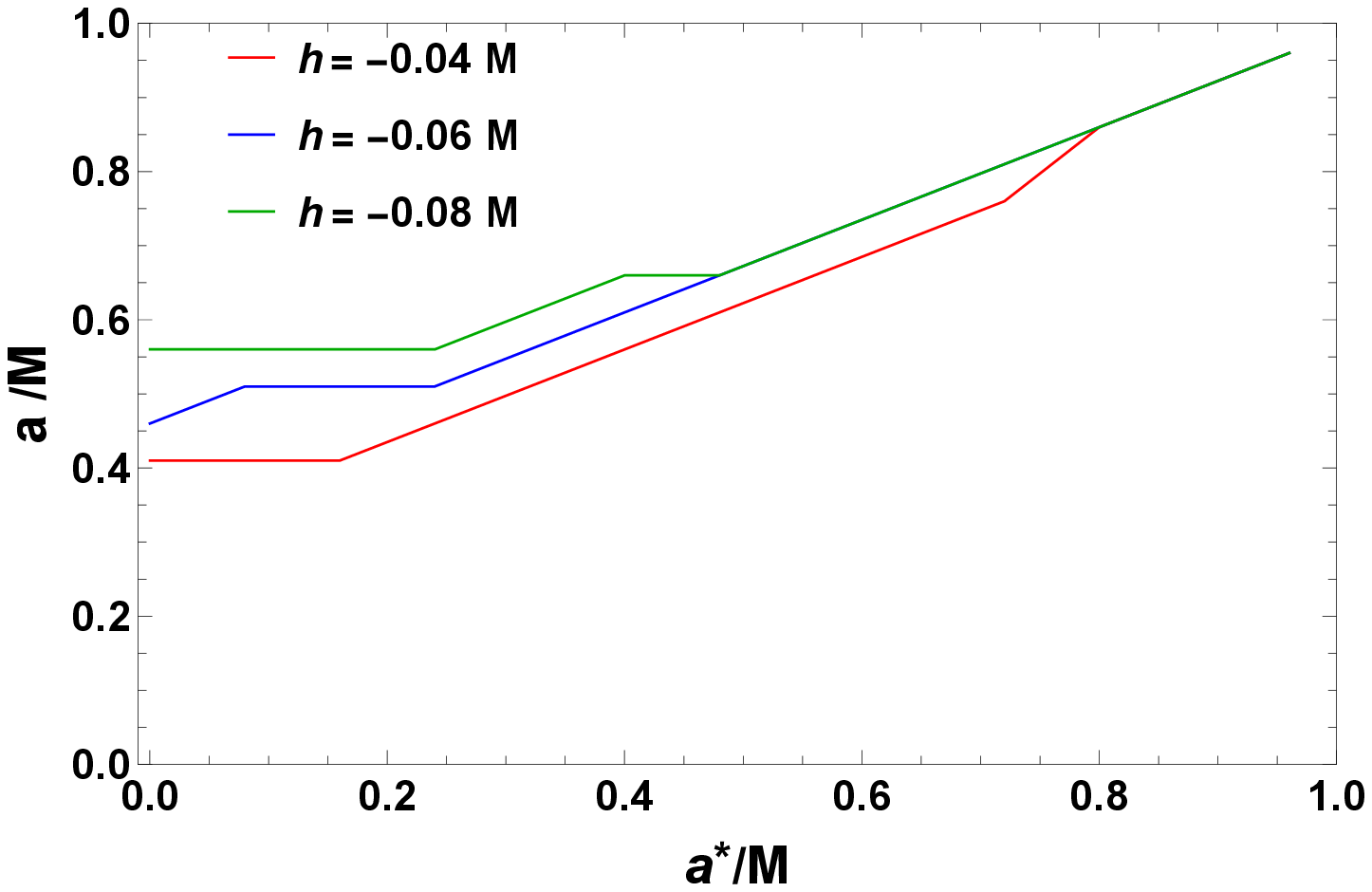}
\end{tabular}
\end{center}
	\caption{The minimized $\chi^2$ (left) and extracted spin $a$ (right) of the best-fit  rotating Horndeski black holes as a function of injected spin $a^*$. The reduced $\chi^2 \leq 1$ for $h\approx-0.04M$.}
	\label{chisquare}
\end{figure*}
\begin{figure*}
\begin{center}
    \begin{tabular}{c c}
    \includegraphics[scale=0.9]{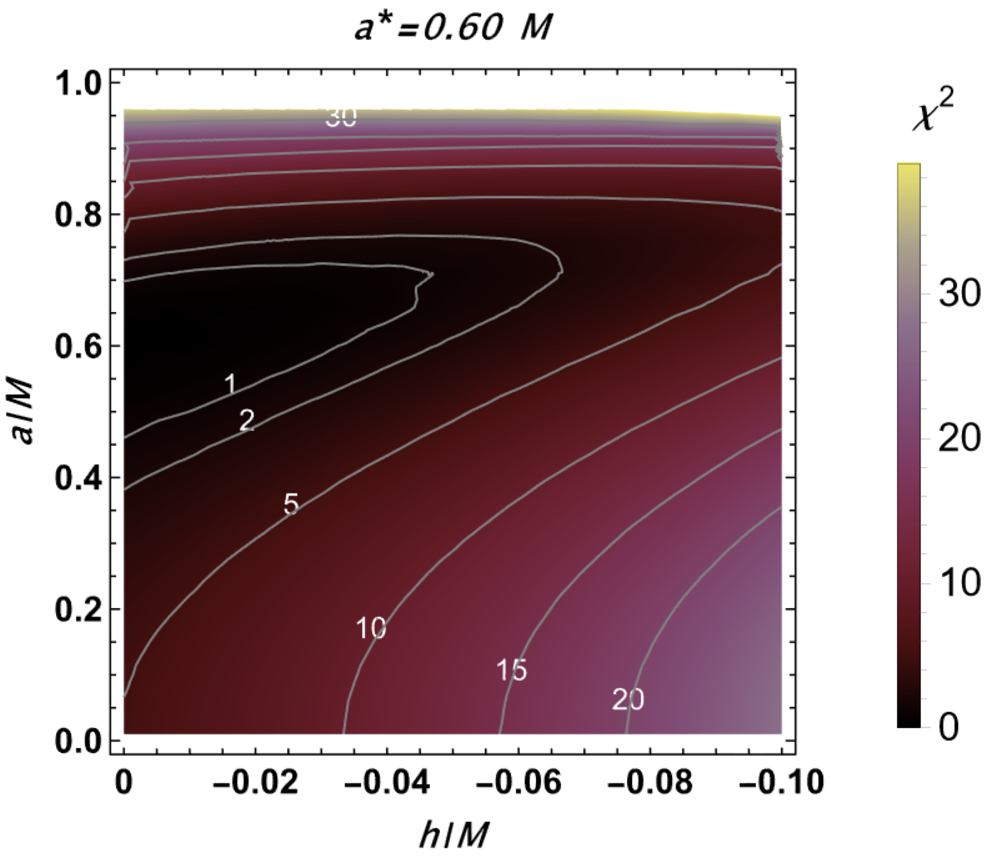}&
     \includegraphics[scale=0.9]{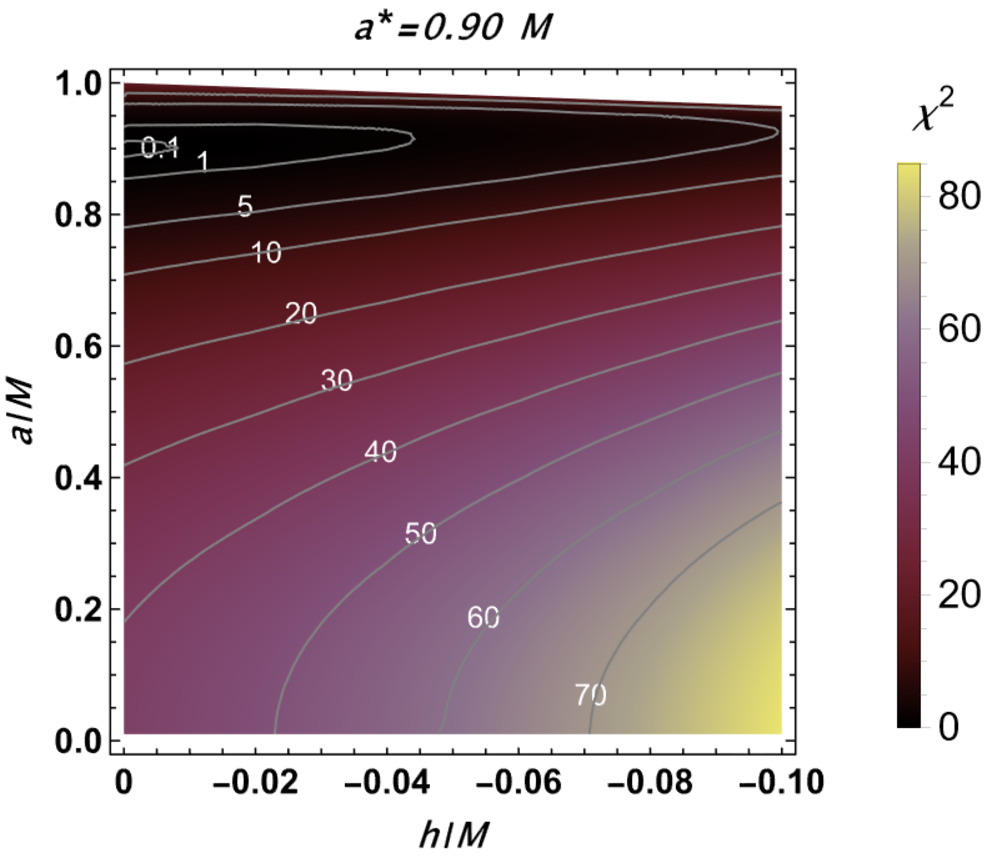}
\end{tabular}
\end{center}
	\caption{The reduced $\chi^2$ between the  rotating Horndeski black hole and the Kerr black hole in the parameter space ($a/M - h/M$) of the former, for the different injected Kerr spin  values $a^*=0.60 M$ (left), $0.90 M$ (right). In the region bounded within the $\chi^2=1$ contours, the  rotating Horndeski black hole cannot be distinguished from the Kerr black hole from the present resolution of the EHT observations.}
	\label{chisquareDensityPlot}
\end{figure*}

We observe from Figure~\ref{shadow2} that the shadows of the Horndeski black holes are different from the Kerr black holes and that the difference is prominent with increasing $|h|$. Indeed, this deviation parameter significantly alters the shape of the Kerr black hole shadow. However, we must check whether these differences are large enough to distinguish the Horndeski gravity from GR with EHT observations, which may require more precise analysis. To probe the possible degeneracy of the rotating Horndeski black hole shadows and the shadows of the Kerr black hole, we conduct a systematic bias analysis. We refer to the Kerr black hole shadow as the injection and the Horndeski black hole shadows as models to fit \citep{Ayzenberg:2018jip,Kumar:2020yem}. It turns out that, for black hole mass $M$ and a fixed observer position ($r_o, \theta_o$), the injected shadow depends on the spin parameter only, whereas model shadows depend on spin $a$ as well as parameter $h$. We adopt the shadow observables $A$ and $D$ from Equations~(\ref{Area}) and (\ref{Oblateness}) to measure the deviation of the model from the injection. We carry out a systematic bias analysis with the reduced $\chi ^2$ merit function between the model and the injection given by Ayzenberg \&Yunes (\citeyear{Ayzenberg:2018jip}) and Kumar et al. (\citeyear{Kumar:2020yem}):
 \begin{equation}
\chi ^2(a,h,a^*)=\frac{1}{2}\sum ^2_{i=1}\bigg[\frac{\alpha ^i(a,h)- \alpha ^i_K(a^*)}{\sigma_i} \bigg]^2,\label{chi_equation}
\end{equation}
where $\alpha^i \equiv \{A,D\}$ are the shadow observables, utilizing 220 sample points \{$h$, $a$\}$\in$\{[$-0.1M$, 0], (0, $a_E$]\}. Here $$\sigma_i=\sqrt{\overline{{\alpha ^i}^2}-{\overline{\alpha ^i}}^2}$$ is the standard deviation in the measurements and $\overline{\alpha ^i}$ denotes the average value of $\alpha ^i$. $\sigma_i$ is taken as $10\%$ of the range of each observable $\alpha^i$, which is the current uncertainty in the observational measurements of the EHT \citep{Akiyama:2019bqs,Akiyama:2019cqa,Akiyama:2019eap,Kumar:2020yem}.
We show that the rotating Horndeski black holes, depending on the values of parameters, in some cases, cast 
shadows that are very similar to those casted by the Kerr black holes ($\chi^2 \le 1$), but in other cases, the two would be clearly distinguishable ($\chi^2 > 1$). 
Thus, for  $\chi ^2\leq 1$ the  rotating Horndeski black hole shadow is degenerate with the Kerr shadow and the two are indistinguishable with the present $10\%$ standard deviation of the EHT observation, whereas $\chi ^2 >1$ signifies that the two shadow observations are astrophysically discernible, and GR can be tested against the Horndeski gravity. Thus we adopt $\chi^2>1$ as a measure of distinguishability of black hole shadows in the two underlying theories of gravity. 

The reduced $\chi^2$ between the model and the injection is minimized, and the corresponding best-fit value of the model spin $a$ is extracted for all injected spin $a^*$.
Figure~\ref{chisquare} shows the extraction results for different values of $h$; the minimized $\chi ^2$ increases with the increase in $|h|$, which is expected, since the $h$ parameter induces deviation from the Kerr shadow (see Figure~\ref{shadow2}). $\chi^2 <1$ for a very small values of $h$, viz., $h=-0.04 M$ (see Figure~\ref{chisquare}), meaning the rotating Horndeski black holes are indistinguishable from the Kerr black holes. Also, for all $h$, $\chi ^2$ increases with the increase in $a^*$ which implies that the shadow distinguishability increases and the near-extremal Kerr black hole can easily be tested against the rotating Horndeski black holes. Furthermore, from the right panel of Figure~\ref{chisquare}, the dependence of best extracted $a$ on $h$ implies that the model (\ref{metric2}), with a higher $|h|$, must spin faster to resemble the injection shadow. Moreover, for higher $a^*$, the best-fit $a$ become degenerate with $h$ (see Figure~\ref{chisquare}).

The Figure~\ref{chisquareDensityPlot} maps the $\chi ^2(a,h,a^*)$ in  the model parameter space for two different injected spins. We find that the region bounded within the $\chi ^2=1$ contour is centered around values of $a$ close to the injected spin $a^*$ and low $|h|$ values, wherein $\chi^2 <1$ and the shadows of the model and the injection degenerate. $\chi^2>1$ is satisfied over a substantial model parameter space, confirming that the shadows of the two black holes are observably different. Indeed, $\chi^2$ increases with increasing $|h|$ (see Figure~\ref{chisquareDensityPlot}). We have examined the behaviour of the $\chi^2(a,h,a^*)$ with $h\in[0,-0.1M]$ to confine within the EHT observational upper bound on $|h|$ (see Figure~\ref{M87obs2}).

Thus, we have explored the possibility of whether the rotating Horndeski and Kerr black holes are astrophysically distinguishable via their shadows within the constrained parameter space consistent with the EHT observations of M87*. We demonstrate that within the accordant parameter space, the reduced $\chi^2$ merit of the the astrophysical observables of the rotating Horndeski and Kerr black holes are large enough to discern the two theories, and it is possible to test GR against the Horndeski gravity. However, our claims may be further strengthened with future astronomical observations like of next-generation EHT \citep{Raymond_2021}, in which the standard deviations in the observables are likely to be less than 10\% ; Equation~(\ref{chi_equation}) would imply that $\chi^2>1$ maybe valid over a larger part of the parameter space ($a$, $h$) than obtained in Figure~\ref{chisquareDensityPlot}, wherein the Horndeski gravity would be distinguishable from its GR counterpart.

The M87* image is also subject to uncertainties arising from various untested accretion and emission models viz., the uncertainties in the plasma physics, coupled with the turbulent nature of the accretion flow. Further, while the EHT observed image of M87* appears to be a ring, it is still ambiguous whether what is seen is actually the lensed image of the unstable photon orbit \citep{Gralla:2020pra,Gralla:2019xty}, the accretion disc/torus around the black hole, or a combination of both. Besides, the stimulated images fitted to the observations have a peak brightness at a radius $\sim$10\% outside the photon ring \citep{Gralla:2019xty,Akiyama:2019bqs}. Here we do not consider the radiative phenomena but compute analytic shadow boundary which represent the innermost region of non-zero emission with a zero flux; this fact alongside the various uncertainties associated with the EHT observations and astrophysical assumptions, means that uncertainties are likely to have been introduce in our investigation---but, with the better resolution images and more certain radiative models of the future, such uncertainties are likely to decrease.
\section{Conclusions}\label{Sec7}
We have constructed the shadow of black holes described by a rotating Horndeski metric, which deviates from the Kerr metric in that it contains an independent $h$, to find that shadow size increases and is more distorted with increasing  $|h|$. 
We have analyzed the various properties of rotating Horndeski black holes with variable parameters ($a$, $h$), estimated the parameters associated with the black hole and put bounds on these parameters by EHT observations. We also analyzed the allowed parameter space for black holes and extremal black holes, and discussed in the detail horizons and the energy emission rate, all of which are critical in the context of astrophysical black holes. Interestingly, the distortion in the rotating Horndeski black hole shadows are also due to the parameter $h$, apart from the spin $a$, and a degeneracy between the shadows of rotating Horndeski black holes with parameter ($a$, $h$) and that of Kerr black hole is observed.
 
 Further, the shadow observables, namely, area $A$ and oblateness $D$, are used to characterize the size and shape of the shadows and, thus, in turn, to estimate the values of black hole parameters. 
We considered the supermassive black holes M87* as rotating Horndeski black hole and used the EHT shadow observables, namely, angular size and asymmetry, to put constraints on the parameter space, viz., $0.0077 M\leq a\leq 0.9353 M$, $-0.7564 M\lesssim h<0$ at $\theta_o=90$\textdegree\, and $0.0048 M\leq a\leq 0.9090 M$, $-0.7920 M\lesssim h<0$ at $\theta_o=17$\textdegree. Thus, within the constrained ($a$-$h$) space, the Horndeski gravity can be used to model M87* within the present observational uncertainties, and subject to the many uncertainties associated with various astrophysical phenomena that obfuscate the EHT measurements. 

As such, we restricted our analysis to the current EHT observational constrained space, and  carried out a systematic bias for distinguishability  of  the rotating Horndeski black holes, to analyze the deviation of their shadows from that of the Kerr black hole. We demonstrate that in some cases the Horndeski black holes, depending on the values of $h$,  produce shadows similar to those produced by the Kerr black hole ($\chi^2 < 1$), but in other cases the two are distinguishable ($\chi^2 > 1$). Indeed, for sufficiently small values of $h$, model shadows significantly differed from the injected shadows, and the current observational facilities can unambiguously discern ($\chi^2 > 1$) the model shadows from the injection shadows. 
 
The GR is a robust theory that has passed all tests exclusively in the weak-field regime \citep{Will:2014kxa}, while the strong-field regime remains practically hardly tested \citep{Psaltis:2008bb}. In future, several observations with unprecedented precision will be available, such as the next-generation EHT \citep{Raymond_2021} and the Event Horizon Imager (EHI) space VLBI array \citep{Roelofs:2021wdi}, opening the door to put further checks on the validity of the no-hair theorem, and, therefore, MoG black holes, like that considered in the present analysis, which may become strong candidates for astrophysical black holes.

\section*{Acknowledgements}
 M.A. is supported by DST-INSPIRE Fellowship, Department of Science and Technology, Govt. of India. S.G.G. thanks SERB-DST for the project No. CRG/2021/005771. 
\bibliography{sample63}{}
\bibliographystyle{aasjournal}

\end{document}